\documentclass[twocolumn]{aastex701}

\usepackage{amsmath}
\usepackage{hyperref}
\usepackage{gensymb}

\usepackage{tikz}
\usetikzlibrary{shapes.geometric, arrows, positioning}
\usepackage{neuralnetwork}
\usepackage{graphicx}

\submitjournal{AAS Journals}
\shorttitle{ULDM \& White Dwarf Structure}
\shortauthors{Crumpler et al.}

\begin{document}

\title{Searching for Ultra-light Dark Matter in Spatial Correlations of White Dwarf Structure}

\author[orcid=0000-0002-8866-4797]{Nicole R. Crumpler}
\email[show]{ncrumpl2@jh.edu}
\altaffiliation{NSF Graduate Research Fellow}
\affiliation{William H. Miller III Department of Physics and Astronomy, Johns Hopkins University, Baltimore, MD 21210, USA}

\author[orcid=0000-0001-6100-6869]{Nadia L. Zakamska}
\email{zakamska@jhu.edu}
\affiliation{William H. Miller III Department of Physics and Astronomy, Johns Hopkins University, Baltimore, MD 21210, USA}

\author[orcid=0000-0002-5864-1332]{Gautham Adamane Pallathadka}
\email{gadaman1@jhu.edu}
\affiliation{William H. Miller III Department of Physics and Astronomy, Johns Hopkins University, Baltimore, MD 21210, USA}

\author[orcid=0000-0002-6871-1752]{Kareem El-Badry}
\email{kelbadry@caltech.edu}
\affiliation{Department of Astronomy, California Institute of Technology, 1200 E. California Blvd., Pasadena, CA 91125, USA}

\date{\today}

\begin{abstract}

If dark matter is ultra-light and has certain Standard Model interactions, it can change the mass-radius relation of white dwarf stars. The coherence length of ultra-light dark matter imparts spatial correlations in deviations from the canonical mass-radius relation, and thus white dwarfs can be used to reconstruct the coherence length, or equivalently the particle mass, of the dark matter field. We simulate the observability of such spatial correlations accounting for realistic complications like variable hydrogen envelope thickness, dust, binaries, measurement noise, and distance uncertainties in DA white dwarfs. Using a machine learning approach on simulated data, we measure the dark matter field coherence length and find that large deviations from the mass-radius relation ($\sim10\%$ change in radius) are needed to produce an observable signal given realistic noise sources. We apply our spatial correlation measurement routine to the SDSS catalog of 10,207 DA white dwarfs. We detect a positive spatial correlation among white dwarfs at separations corresponding to a coherence length of $300\pm50$ pc, with an average Z-score of 85 for white dwarfs separated by less than this coherence length. We conclude that this signal is due to observational bias. The signal can be explained by an offset between measurements and theory for nearby cool white dwarfs, and the presence of few, low-temperature white dwarfs with noisy measurements at further distances. With future improvements in white dwarf models and measurement techniques, particularly for cool white dwarfs, this method can provide interesting constraints on ultra-light dark matter models.

\end{abstract} 


\section{Introduction} \label{sec:intro}

\indent All stars which have initial masses ranging from $\sim0.07-8$ $M_\odot$ end their lives as white dwarfs (WDs) \citep{Fontaine_2001}. Around $80$\% of observed WDs in magnitude-limited samples \citep{Kepler_2019} are DA WDs, WDs with hydrogen-dominated atmospheres showing Balmer series absorption lines in their spectra. DA WDs are the best-understood spectral class of WD, and state-of-the-art models for DA WD spectra are publicly available \citep{Tremblay_2013}. Spectroscopic observations of DA WDs can be used to measure the star's apparent radial velocity, effective temperature ($T_{\text{eff}}$), and surface gravity ($\log{g}$) through comparisons to theoretical spectra. Additionally, photometric observations of WDs can be combined with parallax measurements to measure the star's radius ($R$) and effective temperature  \citep{Bergeron_2019}. 

\indent For typical WD masses of $0.4 - 0.8$ $M_\odot$, the central densities of these stars range from $10^6-10^7$ g cm$^{-3}$ \citep{Boshkayev_2016}. At these densities, WDs are supported against self-gravity largely by electron degeneracy pressure. For low-mass WDs ($\lesssim1 M_\odot$) in the zero-temperature approximation, the resulting mass-radius relation is
\begin{equation}\label{eqn:m_low}
    \frac{M}{M_\odot}=2.08\times10^{-6}\left(\frac{2}{\mu_e}\right)^5 \left(\frac{R}{R_\odot}\right)^{-3},
\end{equation}
where $M$ is the WD mass and $\mu_e$ is the mean molecular weight per free electron. In realistic models of WDs, this simple mass-radius relation is modified by relativistic effects, temperature-dependence, hydrogen envelope thickness considerations, and more \citep{Fontaine_2001}. At a given mass, observed and theoretical WD radii typically agree to within 25\% \citep{Althaus_2013, Camisassa_2016, Camisassa_2019, Crumpler_2025}. Thus, WDs obey a well-defined mass-radius relation that is predominantly determined by electron degeneracy pressure. This relationship has now been well-measured using several techniques to an accuracy comparable to that needed to discover smaller effects, such as temperature-dependence \citep{Vauclair_1997, Bedard_2017, Tremblay_2017, Genest_2019, Chandra_2020_1, Crumpler_2024}.

\indent There has long been convincing observational evidence for the presence of significant amounts of non-baryonic mass in our universe, known as dark matter \citep{Zwicky_1933,Rubin_1970,Riess_1998,Bernardis_2000,Freedman_2001,Planck_2020}. The WDs in our galaxy are embedded in the Galactic dark matter halo, and, due to the relative simplicity of the fundamental physics underlying their structure, have already proven to be an important source of constraints on the masses and interaction strengths for various models of dark matter. Many studies have investigated the cooling or heating effects of different types of dark matter on WDs, and how these effects might impact the WD luminosity function \citep{Raffelt_1986, Isern_2008,Isern_2010, Althaus_2011, Dreiner_2013, Isern_2018} or the pulsational periods of variable WDs \citep{Isern_1992, Corsico_2001, Benvenuto_2004, Biesiada_2004, Bischoff_2008, Corsico_2012_1, Corsico_2012_2}. Other studies have characterized how dark matter might trigger a type Ia supernova, and placed constraints using the observed frequency of these explosions \citep{Graham_2015, Bramante_2015, Graham_2018, Acevedo_2019}. Further approaches to using WDs in dark matter searches include studying the effects of axions on the WD mass-radius relation \citep{Balkin_2024}, investigating how dark matter capture might form small black holes \citep{Steigerwald_2022} or compact cores \citep{Leung_2013} in the centers of WDs, placing constraints dark matter cross sections and densities using very cold WDs in globular clusters \citep{Bertone_2008, McCullough_2010, Bell_2021}, and many more. 

\indent Ultra-light dark matter (ULDM) is a class of dark matter models in which dark matter is comprised of bosons with masses ranging from $m_\text{DM}\sim10^{-24}$ eV to $m_\text{DM}\sim1$ eV, with the precise bounds depending on the details of the model \citep{Ferreira_2021}. Fuzzy dark matter \citep{Hu_2000}, axions and axion-like particles \citep{Choi_2021}, self-interacting dark matter \citep{Spergel_2000}, and dark matter superfluid \citep{Silverman_2002} models are all examples of ULDM. At such low particle masses, ULDM displays wavelike behavior on galactic scales and can be described by a classical field oscillating in time and space \citep{Kaplan_2022}. This wavelike nature results in a rich phenomenology that is distinct from more commonly assumed cold dark matter models with particle masses $\gtrsim$ GeV. On large scales, cosmic microwave background observables and large scale structure surveys agree well with cold dark matter models. However, on sub-galactic scales ($\lesssim 10$ kpc), predictions of cold dark matter models are somewhat inconsistent with observations \citep{Hui_2017}. In particular, the missing satellites \citep{Klypin_1999,Moore_1999}, cusp-core \citep{Flores_1994,Moore_1994}, and too-big-to-fail \citep{Boylan_2011} problems are all known tensions between observations and cold dark matter simulations that may be resolved if dark matter is ultra-light. 

\indent The expectation for a wave-like behavior of ULDM on macroscopic scales enables a variety of constraints on such models \citep{Ferreira_2021}. In particular, the Cosmic Microwave Background restricts ULDM models to $m_\text{DM}\gtrsim10^{-24}$ eV \citep{Hlovzek_2018}, while other constraints challenge the validity of ULDM models with $10^{-24}\lesssim m_\text{DM}\lesssim10^{-21}$ eV \citep{Nori_2019,Schutz_2020,Bar_2022}. Some constraints suffer from uncertainties in determining the properties of small structures with simulations \citep{Zhang_2018} and  weaken if the ULDM model is allowed to have other interactions in addition to its gravitational influence \citep{Ferreira_2021}.

\indent Such interactions can arise through couplings of the ULDM field to ordinary matter. \citet{Damour_2010} consider a variety of couplings between the Standard Model and ULDM, and found that these couplings can generically produce fluctuations in fundamental constants tied to the temporal and spatial variations of the ULDM field. In particular, couplings to the quantum electrodynamics (QED) sector can generate fluctuations in the electromagnetic coupling constant, the muon mass, and the electron mass, and couplings to the quantum chromodynamics sector can change quark masses, which translates into changing the masses of protons and neutrons. A variety of studies have used these variations in fundamental constants to look for ULDM by searching for axion or passing bosonic star effects on terrestrial atomic clocks \citep{Krauss_2019}, by looking for ULDM-induced time modulations in a gravitational wave detector \citep{Vermeulen_2021}, by characterizing the effect of ULDM-induced mass fluctuations on binary pulsars \citep{Blas_2020}, and by looking for spin fluctuations and timing residuals in pulsar timing arrays due to temporal fluctuations from ULDM \citep{Kaplan_2022}. Generally, the couplings considered by \citet{Damour_2010} can also be constrained by searching for fifth forces and equivalence principal violation in laboratory settings in addition to searching for variations in fundamental constants \citep{Damour_2011, Gue_2024}. Because ULDM behaves as a classically oscillating field, Standard Model particles interact with the wave-like nature, not particle-like nature, of ULDM. So, particle-like interactions between ordinary matter and ULDM, such as scattering, are negligible under the couplings considered by \citet{Damour_2010}. 

\indent In this work, we present and carry out a new test of ULDM by characterizing the observability of spatial fluctuations in fundamental constants and other spatially-dependent signatures of ULDM using DA WDs. We show that the wave-like nature of an ULDM model with Standard Model interactions imparts spatial correlations in deviations of the measured masses and radii of nearby DA WDs from the expected mass-radius relation in the absence of ULDM. 

\indent As a motivating example, consider the impact on the electron mass and the structure of WD stars from a QED coupling between ULDM and the Standard Model of the form 
\begin{equation}
    \mathcal{L}_{\phi,\text{ QED}}\supset\frac{\phi}{\Lambda}\left(\frac{d_\gamma}{4e^2}F^{\mu\nu}F_{\mu\nu}\right)-d_em_e\bar{e}e-d_\mu m_\mu\bar{\mu}\mu,
\end{equation}
where $\phi$ is an ultralight scalar field, $\Lambda=M_\text{Pl}/4\pi$, $M_\text{Pl}$ is the Planck mass, and each of $d_\gamma$, $d_e$, and $d_\mu$ are dimensionless coupling coefficients \citep{Damour_2010}. For a review of recent constraints on these couplings, see Fig. 2 of \citet{Vermeulen_2021}. Note that the energy density of the field is $\rho\sim\phi^2$. This coupling would induce a spatial variation in the electron mass of 
\begin{equation}\label{eqn:me}
    m_e(\phi)=(1+\delta m_e)m_{e,\text{ Earth}},
\end{equation}
where $\delta m_e=d_e\phi/\Lambda$ \citep{Kaplan_2022}. Eqn. \ref{eqn:me} assumes a normalization such that the electron mass fluctuates about the Earth electron mass. Allowing for a variable electron mass, we follow the classic derivation of the zero-temperature non-relativistic and relativistic WD mass-radius relations \citep{Chandrasekhar_1933, Stellar_interiors}. We find that, in the non-relativistic limit, the result is a slightly modified version of Eqn. \ref{eqn:m_low}, 
\begin{equation}\label{eqn:RNR}
    \frac{M}{M_\odot}=2.08\times10^{-6}(1+\delta m_e)^{-3}\left(\frac{R}{R_\odot}\right)^{-3}.
\end{equation}
In the relativistic limit, the zero-temperature Chandrasekhar mass is unchanged. Given the non-relativistic and relativistic mass-radius relations, we can interpolate analytically between the two regimes via the \citet{Paczynski_1983} approximation:
\begin{equation}
    R=R_\text{NR}\left[1-\left(\frac{M}{M_\text{C}}\right)^{4/3}\right]^{1/2},
\end{equation}
where $R_\text{NR}$ is the non-relativistic radius from Eqn. \ref{eqn:RNR} and $M_\text{C}$ is the Chandrasekhar mass. Thus, in the presence of an electron mass variation due to an UDLM field, the observed radius of a WD of mass $M$ is modified by
\begin{equation}
    R(\delta m_e)=(1+\delta m_e)^{-1}R_0,
\end{equation}
where $R_0$ is the expected WD radius given the Earth electron mass. The effect of varying the electron mass is to tilt the well-known WD mass-radius relation up or down, leaving the Chandrasekhar mass unchanged. In the case of small deviations from the Earth electron mass ($\delta m_e \ll 1$), the variation in the radius takes the form $R(\delta m_e)=(1-\delta m_e)R_0$. 

\indent We limit the specific case of electron mass spatial variation to a motivating example because there are other astrophysical probes that are potentially better-suited to constrain the variation of the electron mass. Most studies have constrained the temporal rather than the spatial variation of this constant. The two methods in principle test the same type of coupling between ULDM and the Standard Model, but the spatial correlations technique we present here is unique in being able to probe the particle mass. Some such studies have imposed constraints by considering the temporal variation of the electron mass across different cosmological epochs, such as by using quasar absorption systems \citep{Barrow_2005}, the cosmic microwave background radiation \citep{Planck_2015}, or far-infrared fine structure and rotational lines from the reionization epoch \citep{Levshakov_2020}. Other studies use terrestrial laboratory experiments to look for minuscule temporal fluctuations in the electron mass using atomic clocks \citep{Sherrill_2023} or optical clocks and cavities \citep{Kennedy_2020}. Studies constraining the spatial variation of the electron mass within the Galaxy have largely been limited to constraints of the electron-to-proton mass ratio rather than constraints on electron mass variation alone. \citet{Levshakov_2013} used radio observations of the Galactic plane to measure the radial velocity offsets between rotational and inversion transition lines, and placed constraints on the electron-to-proton mass ratio using these observations. Other studies have employed similar approaches, using astrophysical observations of spectral lines to constrain the spatial variation of ratios involving the electron mass, proton mass, and fine structure constant \citep{Levshakov_2008, Levshakov_2010, Vorotyntseva_2024}. Although none of these studies constrain electron mass variation alone, observations of spectral lines are particularly promising sources of constraints on the spatial variation of the electron mass in the Galaxy.

\indent Motivated by the impact of a varying electron mass on the WD mass-radius relation, we consider the effects of generic ULDM-induced variations in WD structure of the form
\begin{equation}\label{eqn:rad_ULDM}
    R(\epsilon)=(1+\epsilon)R_0,
\end{equation}
for $|\epsilon| \ll 1$ where $\epsilon$ depends linearly on the value of the ULDM scalar field and $R_0$ can be found from theoretical mass-radius relations, which include higher order effects on WD structure. Because the ULDM behaves as a coherently oscillating scalar field within the Galaxy, these variations depend on time and the spatial position of the WD.

\indent Nearby DA WDs are typically separated by distances of $10-1000$ pc, meaning spatial correlations of WDs can probe ULDM field coherence lengths on these scales. The  spatial ($\Delta x$) and temporal ($\Delta t$) coherence scales for ULDM field fluctuations come from the wave nature of ULDM, and are given by
\begin{align}
    \Delta x&\sim\frac{\hbar}{\Delta p}\sim\frac{\hbar}{m_\text{DM}v}\\
    \Delta t&\sim\frac{\hbar}{\Delta E}\sim\frac{\hbar}{m_\text{DM}v^2},
\end{align}
where $v$ is the local velocity dispersion of the ULDM. At solar system distances from the Galactic center $v\sim 10^{-3}c$ \citep{Bovy_2012}, and so coherence lengths of $10-1000$ pc correspond to dark matter masses of $m_\text{DM}\sim 6\times10^{-24}-6\times10^{-22}$ eV and timescales of $10^4-10^6$ years. Thus, the spatial distribution of DA WDs can probe ULDM coherence length scales and masses across two orders of magnitude, and the timescales of WD structure variation are sufficiently long that any temporal variations induced by ULDM can be neglected. 

\indent In this paper, we investigate how ULDM can create spatially-correlated deviations between observed and theoretical DA WD masses and radii. When two WDs are within the same ULDM interference patch, with size characterized by the field’s coherence length $\Delta x$, ULDM changes the structure of both stars in a correlated way. Thus, the correlation length of the deviations in WD structure corresponds to the coherence length of the background ULDM field.

In Sec. \ref{sec:catalog}, we briefly describe the catalog of DA WD spectroscopic and photometric parameters used in this work. In Sec. \ref{sec:simulation}, we detail the effects included in our Monte Carlo simulation, which demonstrates how ULDM can impart a spatially-dependent signature on WD structure even in the presence of various sources of noise. We recount our process of training a neural network to reconstruct the ULDM field coherence length from simulations in Sec. \ref{sec:recon_corr_len}. We then apply these methods to search for this signal of ULDM in real data and discuss how this signal can be created in the absence of ULDM in Sec. \ref{sec:data}. We conclude in Sec. \ref{sec:conc}. All spectra are on the vacuum wavelength scale. Surface gravities are measured on the $\log{g}$ scale in dex where $g$, the surface gravity, is in CGS units. The code used for all measurements and results found in this paper is publicly available\footnote{\url{https://github.com/nicolecrumpler0230/ULDM_WD}}.

\section{Catalog} \label{sec:catalog}

\indent We use measurements of DA WD radii, surface gravities, and effective temperatures from the \citet{Crumpler_2025} catalog\footnote{\url {https://www.sdss.org/dr19/data_access/value-added-catalogs/?vac_id=10008}}, which is the largest catalog of both spectroscopic and photometric physical parameters of DA white dwarfs available to date. The catalog contains 8,545 and 19,257 unique DA WDs observed in the 19th Data Release of SDSS-V and previous SDSS data releases, respectively.

\indent The SDSS-V portion of the catalog is comprised of all SDSS-V Data Release 19 DA WDs identified by the spectral classification algorithm \texttt{SnowWhite} through November 2023. The SDSS-V survey \citep{Kollmeier_2025} began operations in November 2020, and operates from the 2.5 m telescopes located at the Apache Point Observatory \citep{Gunn_2006} and Las Campanas Observatory \citep{Bowen_1973}. The Milky Way Mapper program focuses on observing millions of stars in the Galaxy with multi-epoch spectroscopy (J. A. Johnson et al. 2025, in preparation), and WDs observed in SDSS-V were targeted through this program. All SDSS-V WD spectra used in this paper were obtained with the Baryon Oscillation Spectroscopic Survey spectrograph \citep[BOSS, ][]{Smee_2013} using the reduction pipeline v6\_1\_3. 

\indent \citet{Gentile_2021} compiled the observations of DA WDs in the previous SDSS portion of the catalog, covering WDs observed in SDSS through Data Release 16\footnote{\url {https://cdsarc.cds.unistra.fr/viz-bin/cat/J/MNRAS/508/3877\#/browse}}. Data releases from previous generations of SDSS only used the 2.5 m telescope at Apache Point Observatory \citep{Gunn_2006}, and Data Releases 1 through 8 used the original SDSS spectrograph \citep{York_2000}.

\indent \citet{Crumpler_2025} obtain spectroscopic surface gravities and temperatures by fitting the shapes of the first six Hydrogen Balmer series lines (H$\alpha$, H$\beta$, H$\gamma$, H$\delta$, H$\epsilon$, H$\zeta$) using a parametric random forest routine built into the publicly available code \texttt{wdtools\footnote{\url {https://wdtools.readthedocs.io/en/latest/}}} \citep{Chandra_2020_2}. They find that their surface gravity and temperature measurements agree to within $0.060$ dex and $2.4$\%, respectively, when compared to previously published SDSS WD catalogs for spectra with SNR $\geq50$. 

\indent The photometric radii and effective temperatures contained in the \citet{Crumpler_2025} catalog are measured by fitting a combination of cross-matched Gaia DR3 photometry \citep{Gaia_2016}, SDSS Data Release 17 photometry \citep{Abdurrouf_2022}, and \citet{BailerJones_2021} distances to model photometry via $\chi^2$ minimization. No WD mass-radius relation is assumed during the fitting process. The observed photometry is corrected for extinction using the three-dimensional dust map of \citet{Edenhofer_2024} from the \texttt{dustmaps}\footnote{\url {https://dustmaps.readthedocs.io/en/latest/index.html}} Python package \citep{Green_2018} and the extinction curve from \citet{Fitzpatrick_1999} from the \texttt{extinction}\footnote{\url {https://extinction.readthedocs.io/en/latest/}} Python package. The model photometry is created by convolving \citet{Tremblay_2013} model spectra through photometric filter response curves from the \texttt{pyphot}\footnote{\url {https://mfouesneau.github.io/pyphot/}} Python package. \citet{Crumpler_2025} find that their radius and temperature measurements agree to within $0.0005$ $R_\odot$ and $3$\%, respectively, when compared to the \citet{Gentile_2021} catalog for fits to Gaia photometry. 

\indent To ensure high-quality measurements, we apply data quality cuts to the catalog to obtain our final sample. Following \citet{Crumpler_2025}, we keep only objects for which the SNR of the coadded spectrum is $>10$. We only use spectroscopic parameters from coadded spectra in this work. The SNR cut is by far the most stringent, and reduces the number of objects in the SDSS-V and previous SDSS catalogs to 3,772 and 9,677, respectively. \citet{Tremblay_2013} model spectra cover a range of $1,500<T_{\text{eff}}<130,000$ K and $7<\log{g}<9$ dex, so we remove any object for which the spectroscopic surface gravity is outside the valid surface gravity range or for which the spectroscopic or photometric temperature is outside the temperature range of $1,600<T_{\text{eff}}<129,000$ K. We also cut on the errors of each measured parameter to ensure that all measurements we use are sufficiently reliable. These cuts restrict the full error on the measured photometric radius to $<0.006$ $R_\odot$ and on the measured spectroscopic surface gravity to $<0.3$ dex. We then remove any objects flagged as potential binaries by \citet{Crumpler_2025}. Finally, we restrict our sample to objects with \citet{BailerJones_2021} median geometric distances $<1,000$ pc. Beyond $\sim1,000$ pc, the number of WD observations declines appreciably, but this is not a significant restriction because there are few high quality spectra of WDs beyond 1,000 pc. After all quality cuts, there are $3,006$ and $7,918$ unique DA WDs remaining in the SDSS-V and previous SDSS samples, respectively. 

\indent After implementing data quality cuts, we combine the SDSS-V and previous SDSS samples into a single catalog. For each measured parameter, we take the weighted mean of the SDSS-V and previous SDSS measurement to obtain a single value. This results in a final sample of 10,207 unique DA WDs with high-quality spectroscopic and photometric measurements within 1000 pc. This catalog of combined SDSS-V and previous SDSS measurements passing quality cuts is hereafter referred to as the clean catalog.

\section{Simulation} \label{sec:simulation}

\indent In this section, we describe how we build a simulated sample of DA WD observations, with various noise sources, overlaid on an idealized ULDM background for different maximum-deviations of the observed WD radius from the expected theoretical radius ($\epsilon_\text{max}$) and ULDM field coherence lengths ($\Delta x$). We convert these simulated WDs into curves characterizing the extent of spatial correlation in WD structure as a function of the separation between two WDs.

\subsection{Including ULDM Effects} \label{sec:ULDM}

\indent Let $\Phi$ be the typical amplitude of the ULDM scalar field. We choose a normalization such that the observed WD radius is the expected theoretical radius ($\epsilon=0$) when the classically oscillating ULDM scalar field is at its typical value ($\phi=\Phi$), the observed radius is at a maximum ($\epsilon=\epsilon_\text{max}$) when the ULDM scalar field is at its maximum ($\phi=2\Phi$), and the observed radius is at a minimum ($\epsilon=-\epsilon_\text{max}$) when the ULDM scalar field is at its minimum ($\phi=0$). Thus, $\epsilon_\text{max}$ represents the maximum deviation of the observed WD radius from the expected theoretical radius of the star.

\begin{figure}[h!]
\begin{center}
\includegraphics[scale=0.45]{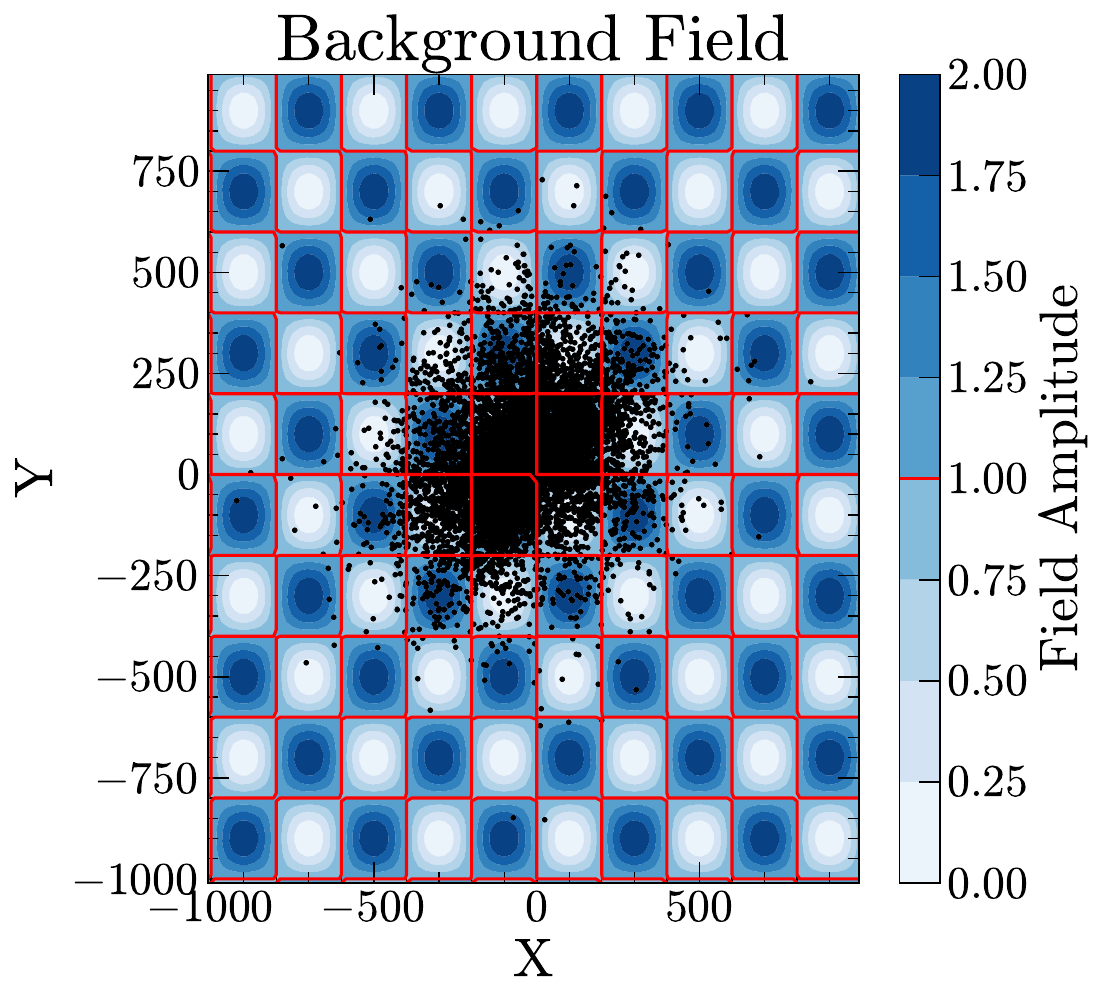}
 \caption{A simulated sample of 10,207 WDs (black circles) with coordinates drawn from the clean catalog of Sec. \ref{sec:catalog}, overlaid on an idealized ULDM background scalar field with coherence length $\Delta x=200$ pc.\label{fig:uldm_field}}
\end{center}
\end{figure}

\indent For simplicity, we model the background ULDM scalar field as a three-dimensional sinusoidal field given by
\begin{equation}\label{eqn:field}
    \phi(U,V,W)=\Phi
    \left[\sin{\left(\frac{\pi U}{\Delta x}\right)}\sin{\left(\frac{\pi V}{\Delta x}\right)}\sin{\left(\frac{\pi W}{\Delta x}\right)}+1\right],
\end{equation}
where $U$, $V$, and $W$ are the Galactic Cartesian coordinates of the point, $\Delta x$ is the coherence length of the field, and $\Phi$ is the typical amplitude of the ULDM scalar field. For comparison, \citet{Yavetz_2022} present numerical solutions of the Schr\"odinger-Poisson equations to show a more realistic rendering of the dark matter density ($\rho\sim\phi^2$) in an ULDM halo. A two-dimensional slice of this simple three-dimensional model for a background field with coherence length $\Delta x=200$ pc is displayed in Fig. \ref{fig:uldm_field}. On the field, we have overlaid a sample of 10,207 WDs with coordinates drawn from the clean catalog of Sec. \ref{sec:catalog}. Most WDs have coordinates within $\pm500$ pc of the origin, so most WD separations are $1000$ pc or nearer.

\indent Given a WD of fixed mass $M_\text{WD}$ and initial, unperturbed radius $R_\text{WD, init}$ at a $(U,V,W)$ coordinate within the ULDM field, the radius and surface gravity of the star are modified according to
\begin{align}
    \epsilon&=\epsilon_\text{max}\sin{\left(\frac{\pi U}{\Delta x}\right)}\sin{\left(\frac{\pi V}{\Delta x}\right)}\sin{\left(\frac{\pi W}{\Delta x}\right)} \notag\\
    R(\epsilon)&=(1+\epsilon)R_\text{WD, init} \label{eqn:uldm_sim}\\
    \log{g}(\epsilon)&=\log{g_\text{WD, init}}-2\log{(1+\epsilon)},\notag
\end{align}
where $g_\text{WD, init}=\frac{G M_\text{WD}}{R_\text{WD, init}^2}$. The effects of ULDM on the WD mass-radius relation are included using these analytical expressions.

\subsection{Calculating Spatial Correlations} \label{sec:moransI}

\indent From Eqn. \ref{eqn:uldm_sim}, if two WDs are within the same field interference peak, the radii of both WDs increase and the surface gravities decrease relative to the expected values from the WD mass-radius relation. Similarly, if two WDs are within the same field interference trough, the radii of both WDs decrease and the surface gravities increase relative to the expected values from the WD mass-radius relation. This spatial correlation of deviations from the WD mass-radius relation can be used to probe the coherence length, and thus the mass, of ULDM models.

\indent In order to characterize this spatial correlation, we employ a statistic known as Moran's I \citep{Moran_1950}. Moran's I is a measure of spatial autocorrelation defined by
\begin{equation}\label{eqn:moransI}
    I=\frac{N}{\mathbb{W}}\left(\sum_i\sum_j w_{ij}(x_i-\bar{x})(x_j-\bar{x})\bigg/\sum_i(x_i-\bar{x})^2\right),
\end{equation}
where $N$ is spatial units indexed by $i$ and $j$, $x$ is the variable of interest, $\bar{x}$ is the expected value of $x$, $w_{ij}$ is the ($i$th, $j$th) element of a matrix of spatial weights with zeroes on the diagonal ($w_{ii}=0$), and $\mathbb{W}$ is the sum of all weights ($\sum_i\sum_j w_{ij}$). The choice of weight matrix varies depending on the application of the statistic. The significance of the extent of spatial autocorrelation is given by the Moran's I Z-score, computed as
\begin{equation}
    Z=\frac{I-E[I]}{\sqrt{V[I]}},
\end{equation}
where $E[I]$ is the expected value and $V[I]$ is the variance of $I$. The expected value of $I$ assuming the null hypothesis of no spatial autocorrelation is given by
\begin{equation}
    E[I]=\frac{-1}{(N-1)}.
\end{equation}
The variance is given by
\begin{align}
    V[I]&=E[I^2]-(E[I])^2\\
    E[I^2]&=\frac{NS_4-S_3S_5}{(N-1)(N-2)(N-3)\mathbb{W}^2}\\
    S_1&=\frac{1}{2}\sum_i\sum_j(w_{ij}+w_{ji})^2\\
    S_2&=\sum_i(\sum_jw_{ij}+\sum_jw_{ji})^2\\
    S_3&=\frac{N^{-1}\sum_i(x_i-\bar{x})^4}{(N^{-1}\sum_i(x_i-\bar{x})^2)^2}\\
    S_4&=(N^2-3N+3)S_1-NS_2+3\mathbb{W}^2\\
    S_5&=(N^2-N)S_1-2NS_2+6\mathbb{W}^2.
\end{align}
Moran's I usually ranges from -1 to +1, with values  below $-1/(N-1)$ indicating negative spatial autocorrelation and values above $-1/(N-1)$ indicating positive spatial autocorrelation.

\indent We are searching for spatial correlations in the difference between the expected radius from the mass-radius relation and the measured radius. We have the measured ($U,V,W$) coordinates, radius, surface gravity, and temperature for each WD. We can then measure the mass of the WD independently of an assumed mass-radius relation using,
\begin{equation}
    M=\frac{gR^2}{G}.
\end{equation}
To obtain the theoretical radius of the WD based on the WD mass-radius relation, we use the La Plata models\footnote{\url {http://evolgroup.fcaglp.unlp.edu.ar/TRACKS/newtables.html}}. These models contain tables of DA WD masses and radii as a function of effective temperature and surface gravity. For low-mass helium core WDs, intermediate-mass carbon-oxygen core WDs, and high-mass oxygen-neon core WDs, these models use the results of \citet{Althaus_2013}, \citet{Camisassa_2016}, and \citet{Camisassa_2019}, respectively. The difference between the theoretical radius and the measured radius is $x$ in Eqn. \ref{eqn:moransI}. In this application, $N$ is the number of WDs in the sample, $\bar{x}$ is 0 since we expect the WDs to follow the mass-radius relation, and we choose a weight matrix such that $w_{ij}=1$ if the distance between two WDs is less than a distance cutoff of $d$ and $w_{ij}=0$ otherwise.

\indent We calculate Moran's I and the Z-Score of the statistic for 46 distance cutoffs between $d=1$ and $d=1000$ pc. At each distance cutoff, $I$ and $Z[I]$ characterize the significance of the positive or negative spatial autocorrelation of the deviation from the WD mass-radius relation. 

\subsection{Initial Parameters} \label{sec:init_param}

\indent The last step necessary to initialize the simulation is to create a realistic sample of simulated WD spatial coordinates, radii, surface gravities, and temperatures in the absence of ULDM. We can apply the analytical expression describing the impact of ULDM on a WD to generate a sample of WDs in an ULDM field. Given this simulated sample of WDs in an ULDM field, we then calculate Moran's I to characterize the spatial correlation imparted by the background field. To create these simulated WD physical parameters for our Monte Carlo simulation, we use the clean catalog of Sec. \ref{sec:catalog}

\indent Astronomical measurements often have  distance-dependent biases. Smaller and dimmer objects are typically difficult to observe at large distances, and measurement errors usually increase as a function of distance. To account for this, we bin the clean catalog of Sec. \ref{sec:catalog} in distance intervals of $50$ pc. We draw a sample of 10,207 coordinates for each WD from the median geometric distances (\texttt{r\_med\_geo}) and Galactic coordinates (\texttt{l}, \texttt{b}) in the clean catalog. We find that the \texttt{l} and \texttt{b} distributions in the catalog are not distance-dependent, so we do not utilize the distance bins in drawing these parameters. We convert these coordinates into Galactic Cartesian coordinates ($U,V,W$) using \texttt{astropy}. Then, we draw a mass and effective temperature for each simulated WD from the catalog objects contained in the corresponding distance bin. To account for correlations between mass and temperature, we draw these two parameters simultaneously. We use masses measured from photometric radii and the theoretical La Plata Models (\texttt{mass\_rad\_theory}) and photometric effective temperatures  (\texttt{teff\_phot}) to create this sample. With these mass and temperature samples, we use the La Plata models to create radius and surface gravity samples which obey the theoretical mass-radius relation. These samples do not yet contain any sources of noise, such as measurement uncertainty. The impact of noise is considered in later sections.

\indent We record the initial mass, temperature, radius, surface gravity, and coordinate samples of these simulated WDs. We then modify both the radius and surface gravity of each simulated WD according to Eqn. \ref{eqn:uldm_sim}, given the coordinates of each WD and the chosen ULDM field parameters $(\Delta x,\epsilon_\text{max})$, to obtain the measured radius and surface gravity. With the measured radius and surface gravity, we compute a measured mass for each WD which includes the effects of ULDM. The theoretical radius for the simulated WD is calculated by combining the measured mass and temperature with the La Plata models. The deviation of the WD from the theoretical mass-radius relation is calculated as the difference between the theoretical radius and the measured radius. 

\begin{figure}[h!]
\begin{center}
\includegraphics[scale=0.25]{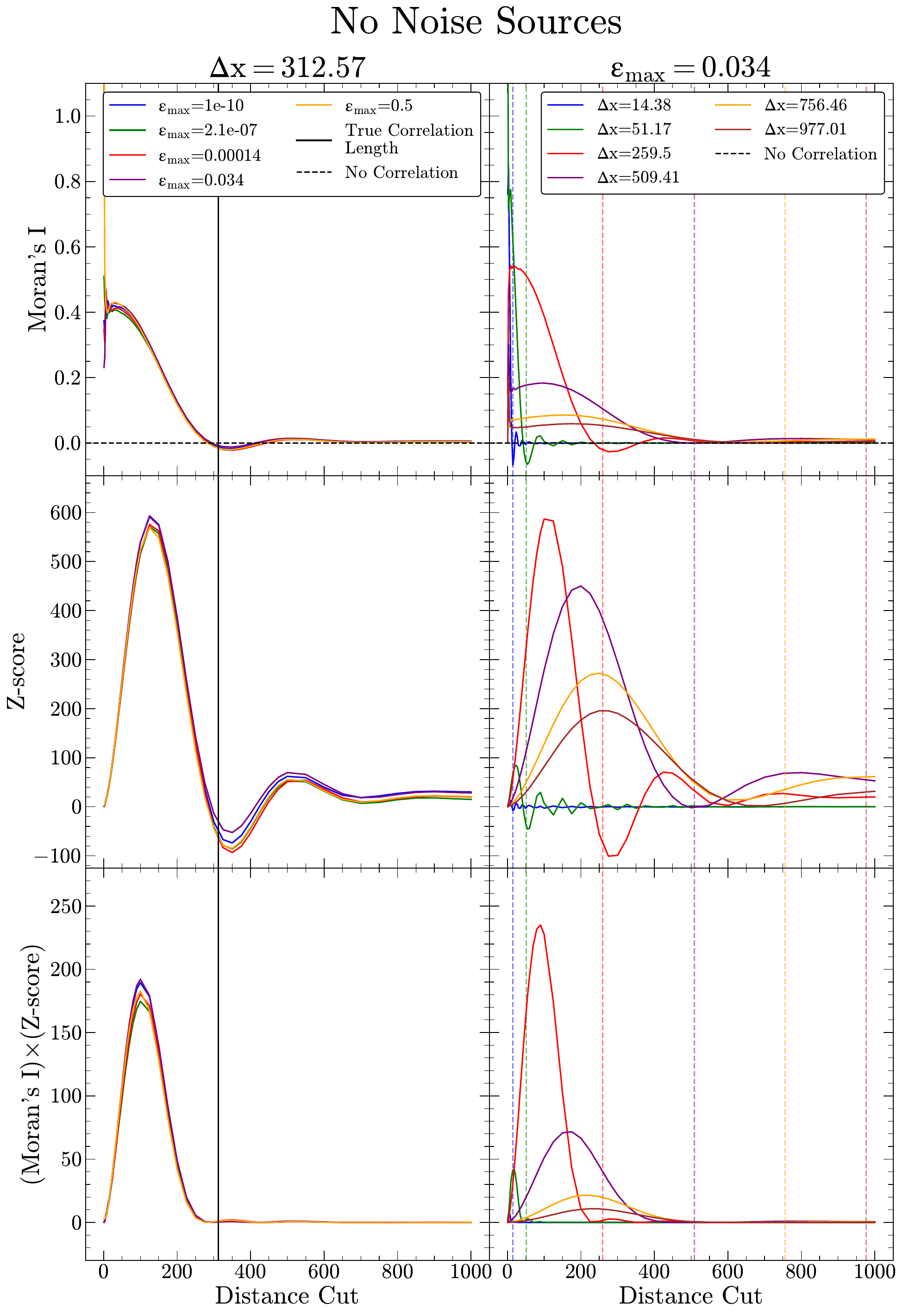}
\caption{The spatial autocorrelation signal from a simulated catalog of 10,207 WDs with no added sources of noise. The top, middle, and bottom panels show the Moran's I statistic, Z-score, and product of Moran's I and Z-score as a function of the maximum distance between two WDs included in the calculation. The left column shows these statistics for one value of the ULDM coherence length and various maximum observed radius deviations, and the right column shows these statistics for a particular maximum observed radius deviation and various ULDM coherence lengths. The shape of the (Moran's I)$\times$(Z-score) curve indicates the coherence length of the field, with the peak occurring at $\sim1/2-1/3$ the coherence length and the curve reaching a minimum at distances just smaller than the coherence length. \label{fig:curves_no_noise}}
\end{center}
\end{figure}

\indent We run a set of $2,921$ simulations for different combinations of $1<\Delta x<1000$ pc and $10^{-10}<\epsilon_\text{max}<0.5$ where we overlay the simulated WDs on an UDLM background and compute the spatial autocorrelation according to Sec. \ref{sec:moransI} with no added sources of noise. Fig. \ref{fig:curves_no_noise} shows the outputs of some of these simulations for different combinations of $\Delta x$ and $\epsilon_\text{max}$. The curves all show the same characteristic behavior. For the Moran's I curve, we see a strong positive spatial correlation at WD separations less than the coherence length, and the strength of this spatial correlation decreases as the WD separations approach the coherence length. In the Z-score curve, the significance of the positive correlation is low at small WD separations because there are fewer WDs at these separations. This significance increases to a maximum at $\sim 1/2 - 1/3$ of the coherence length and then begins to decline again. The decline in both $I$ and the Z-score at separations approaching $\Delta x$ arises because even if two WDs are separated by less than $\Delta x$, they may not fall within the same interference patch of the ULDM field. As WD separations approach the coherence length, the likelihood of two WDs falling within the same peak or trough of the field decreases and the strength and significance of the positive correlation likewise decrease. In Fig. \ref{fig:curves_no_noise}, the ULDM signal is visible for all simulated maximum deviations of the electron mass ($\epsilon_\text{max} \geq 10^{-10}$). Although increasing $\epsilon_\text{max}$ increases the amplitude of the deviations, Moran's I is dimensionless. Thus, in the absence of noise sources, the extent and significance of the spatial correlations do not increase. The value of $\epsilon_\text{max}$ becomes important in the presence of noise sources, when the amplitude of the deviations must be sufficiently large to overcome variations due to noise.

\indent We repeat these simulations including various sources of noise and contamination that are typical in realistic DA WD observations. Each of these noises sources are described in the subsequent sections.

\subsection{Including Variable Hydrogen Envelope Thickness} \label{sec:thinH}

\indent There is evidence that the thickness of the hydrogen layer surrounding DA WDs varies by orders of magnitude, with some WDs having thick hydrogen envelopes \citep[$M_\text{H}/M_\text{WD} \sim 10^{-4}$, ][]{Iben_1984} and others having thin envelopes \citep[$M_\text{H}/M_\text{WD} \sim 10^{-10}$, ][]{Fontaine_2001}. Although the number of thin layer DA WDs is thought to be far less than those with thick hydrogen layers \citep[$\lesssim20$\%, ][]{Tremblay_2008}, the hydrogen layer thickness has  important consequences for measuring the radius and surface gravity of DA WDs \citep{Crumpler_2024}. Both the \citet{Tremblay_2013} and La Plata models employed in the \citet{Crumpler_2025} catalog and this paper utilize thick hydrogen layers. Thin hydrogen layer WDs have smaller radii by $\sim 0.0006$ $R_\odot$ when compared to thick hydrogen layer WDs of the same mass \citep{Crumpler_2024}.

\indent We simulate the impact of thin hydrogen layer WD contamination on our ability to detect the ULDM signal described in Sec. \ref{sec:ULDM}. We repeat the steps of Sec. \ref{sec:init_param}, drawing an initial sample of WD parameters with no sources of noise. We  modify the observed radius and surface gravity of each simulated WD according to Sec. \ref{sec:ULDM}. Then, as conservative estimate, we assume the highest contamination fraction indicated by the literature, and randomly select 20\% of the simulated sample to be thin hydrogen layer WDs. For each selected WD, we adjust the measured surface gravity and radius to values assuming the object instead has a thin hydrogen layer. We obtain these adjusted values by taking the difference in the radius or surface gravity predicted by the thin and thick hydrogen layer models of \citet{Fontaine_2001} at the measured WD photometric temperature and mass. We then compute the spatial correlation as in Sec. \ref{sec:moransI} for various $1 \leq\Delta x \leq 1000$ pc and $10^{-10} \leq  \epsilon_\text{max} \leq 0.5$. We find that thin hydrogen layer WD contamination has little impact on the ULDM signal, and the ULDM signal is visible for all simulated maximum deviations of the electron mass ($\epsilon_\text{max} \geq 10^{-10}$). This may be because thin hydrogen layer WDs still abide by a physical mass-radius relation, although it is shifted slightly from the mass-radius relation for WDs with thick hydrogen layers.

\subsection{Including Dust Effects} \label{sec:dust}

\indent Dust imparts spatially-dependent signals on measured WD photometric radii and effective temperatures. These signals can interfere with our ability to detect the ULDM signal described in Sec. \ref{sec:ULDM}. To investigate the impact of extinction, we simulate the effect of dust on spatial correlations of WD structure. First, we re-measure all photometric radii and effective temperatures for each WD included in the clean catalog of Sec. \ref{sec:catalog} following the measurement procedures of \citet{Crumpler_2025} without including any extinction corrections. The \citet{Crumpler_2025} catalog already contains photometric measurements corrected for extinction. We then compare the measured photometric parameters for all objects in the catalog with and without extinction corrections. We bin all measurements in the Johnson V-band extinction ($\text{A}_\text{V}$) and take the median difference between the extinction-corrected and uncorrected measured radii and temperatures for each bin. Thus we obtain, as a function of $\text{A}_\text{V}$, the impact of dust on the measured photometric parameters. We find that radii measured without extinction corrections are larger and temperatures are cooler compared to extinction-corrected values.

\begin{figure}[h!]
\begin{center}
\includegraphics[scale=0.25]{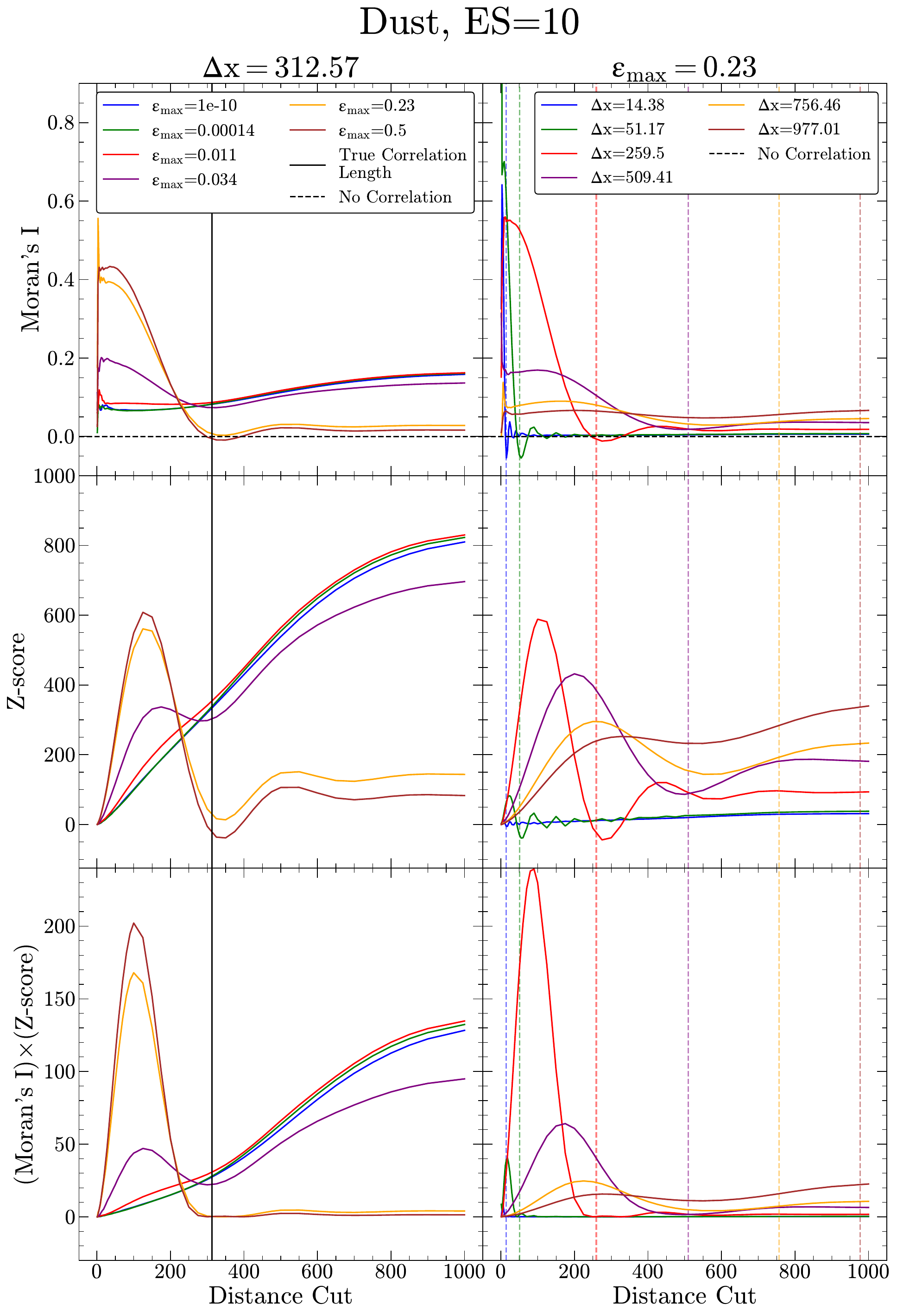}
\caption{Same as Fig. \ref{fig:curves_no_noise}, but with noise due to dust added. We reduce the strength of the dust effect by a factor of 10. For sufficiently small $\epsilon_\text{max}$, the ULDM signal is washed out by the effect of dust, resulting in a monotonic increase of the (Moran's I)$\times$(Z-score) curve with increasing distance cutoff. \label{fig:curves_dust_10}}
\end{center}
\end{figure}

\indent We repeat the steps of Sec. \ref{sec:init_param}, drawing an initial sample of WD parameters with no sources of noise. We modify the simulated radii and surface gravities according to Sec. \ref{sec:ULDM}. Then, for each simulated WD we obtain the $\text{A}_\text{V}$ extinction at the three-dimensional coordinates of the WD from the \citet{Edenhofer_2024} dust map. We identify the $\text{A}_\text{V}$ bin and corresponding median difference between the extinction-corrected and uncorrected radius and temperature for that WD. To tune the strength of the effect of dust on the spatial correlations of WD deviations from the mass-radius relation, we divide the median difference between the extinction-corrected and uncorrected radius and temperature by an \texttt{effect\_strength} parameter. We adjust the WD photometric parameters by subtracting the median difference for that $\text{A}_\text{V}$ bin divided by the \texttt{effect\_strength} from the initial WD radius and temperature. For objects nearer than 69 pc, we set $\text{A}_\text{V}=0$ and do not change the simulated photometric measurements. We then compute the spatial correlation as in Sec. \ref{sec:moransI} for various $1 \leq\Delta x \leq 1000$ pc and $10^{-10} \leq  \epsilon_\text{max} \leq 0.5$. Fig. \ref{fig:curves_dust_10} shows the outputs of some of these simulations with only dust effects for different combinations of $\Delta x$ and $\epsilon_\text{max}$. In Fig. \ref{fig:curves_dust_10}, we reduce the impact of dust by an order of magnitude by setting \texttt{effect\_strength}=10. Without reducing the impact of dust, the simulation would assume that the measured WD photometric parameters are completely uncorrected for dust. Reducing the effect of dust simulates the case in which some, but not all, of the effects of dust are accounted for in the \citet{Edenhofer_2024} dust map.

\indent We find that for $\epsilon\lesssim0.034$, the ULDM signal is no longer observable due to the large deviations from the mass-radius relation caused by dust. Decreasing the \texttt{effect\_strength} parameter causes the ULDM signal to disappear at larger $\epsilon$. For \texttt{effect\_strength}=1, the ULDM signal is only observable for $\epsilon\gtrsim 0.1$, where the exact cutoff also depends on the coherence length of the field. When the ULDM signal is no longer observable the presence of dust creates a positive correlation on all spatial scales. The Z-score of this positive correlation becomes very large on long distance scales, causing the (Moran's I)$\times$(Z-Score) curve to be monotonically increasing. 

\subsection{Including Binary Contamination} \label{sec:binary}

\indent The \citet{Crumpler_2025} catalog includes a flag to remove double WD (DWD) binary contaminants, for which the measured radius, surface gravity, and temperature contained in the catalog are incorrect. In Sec. \ref{sec:catalog}, we remove objects flagged as binaries from our clean catalog, however, this binary flag is simplistic and likely misses more subtle binaries in the catalog. We simulate how these remaining binary contaminants impact our ability to detect the effect of ULDM on spatially correlated deviations from the WD mass-radius relation. First, we characterize how binaries impact the measured parameters of DA WDs by creating two simulated WD samples, one with no binaries and one only with binaries, to investigate how, on average, binaries affect the measured WD radius, surface gravity, and temperature. We create a no-binary sample of 10,000 simulated WDs with masses and temperatures drawn from the \texttt{mass\_rad\_theory} and \texttt{teff\_phot} columns, respectively, of the clean catalog of Sec. \ref{sec:catalog}.  As in Sec. \ref{sec:init_param}, we obtain the radii and surface gravities for this no-binary sample.

\indent Then we create a corresponding DWD binary sample in which we assume every star is a binary system. For each system, we draw a companion mass from our primary mass sample, requiring the companion to be more massive and thus smaller than the primary WD. We also draw a temperature from our primary temperature sample and compute the companion radius and surface gravity. We draw a system orbital separation from the \citet{Maoz_2018} distribution, and use it to compute the period of the system as well as the orbital velocities of both the primary and companion assuming circular orbits. Following \citet{Maoz_2012}, we draw a line-of-sight inclination from $P(i)\propto \sin(i)$, choose a random orbital phase, and calculate the primary and companion radial velocities. We then use spectral templates by \citet{Tremblay_2013} to build model spectra for both the primary and companion WD in each binary system. We add these spectra to obtain the model spectrum for the full binary system. Also, we obtain model SDSS \textit{urz} and Gaia $G_\text{BP}$ and $G_\text{RP}$ photometry for the primary and companion using the \citet{Fontaine_2001} interpolation from mass and effective temperature to absolute magnitudes implemented in the publicly available package \texttt{WDmodels}\footnote{\url {https://github.com/SihaoCheng/WD_models}}. We add these model magnitudes in flux space and convert them back to magnitudes to obtain the SDSS or Gaia photometry for the whole binary system. We then repeat the measurement procedures of \citet{Crumpler_2025} to obtain spectroscopic and photometric parameters in the case when each WD is isolated and each WD is in a DWD binary. 

\indent We compare the measured photometric and spectroscopic parameters for all simulated objects with and without binaries. Separately for Gaia and SDSS photometry, we bin the difference between the primary WD photometric radius and DWD binary system radius in primary radius,  and calculate the median difference in each bin to characterize the typical impact of binarity given the radius of the WD. For both SDSS and Gaia photometry, the presence of binaries results in larger measured radii for primaries with radii $\lesssim 0.02-0.025 R_\odot$, while the opposite trend holds for primaries with larger radii. We repeat this process for photometric effective temperatures to also obtain the typical effect of binaries as a function of the temperature of the primary star. For SDSS photometry, binaries result in hotter apparent effective temperatures for primaries with temperatures $\lesssim 20,000$ K and cooler apparent temperatures for hotter primaries, while for Gaia photometry binaries always result in hotter median apparent effective temperatures.  For spectroscopic surface gravities, we bin the difference between the primary WD surface gravity and DWD binary system value in primary surface gravity, and calculate the median difference in each bin to characterize the typical impact of binarity given the surface gravity of the WD. For WDs with primary surface gravities $\lesssim 8.1$ dex, a binary companion results in a larger measured surface gravity than that of the primary alone. For WDs with larger surface gravities, the companion causes a decrease in the apparent surface gravity. 

\indent Given the typical effect of binarity as a function of radius, temperature, or surface gravity, we can simulate how noise from binary contamination can impact the signal of ULDM. We repeat the steps of Sec. \ref{sec:init_param}, drawing an initial sample of WD parameters with no sources of noise. We modify the simulated radii and surface gravities according to Sec. \ref{sec:ULDM}. Then, we add DWD binary contaminants. For each WD, if the WD mass is $<0.45~M_\odot$, we set the probability of having a binary companion equal to 1 \citep{Marsh_1995}. At such low masses, WDs must have evolved via binary evolution in order to become a WD over the age of the Universe. If the primary mass is greater than this threshold, we give the star a 1\% chance of having a binary companion. One percent is chosen because the unresolved WD binary percentage found in the literature varies from 1\%–10\% \citep{Holberg_2009,Toonen_2017,Maoz_2018,Torres_2022}. Overall, this results in a binary sample with a total binary proportion of $\sim8-10$\%, which is in line with expectations from the literature and is a conservative choice considering we remove objects flagged as binaries by \citet{Crumpler_2025}. We identify the radius, temperature, and surface gravity bins for each DWD binary and the corresponding median difference between the binary and no-binary measurements. We adjust the WD radius, temperature, and surface gravity by subtracting the median difference for that bin from the measured WD value. For the photometric parameters, we take the average of the difference with and without binaries for SDSS and for Gaia photometry. We then compute the spatial correlation as in Sec. \ref{sec:moransI} for $1 \leq\Delta x \leq 1000$ pc and $10^{-10} \leq  \epsilon_\text{max} \leq 0.5$. We remove any objects with observed radii $>0.016~R_\odot$ from the simulated sample since these are low-mass stars that are likely binaries \citep{Marsh_1995}. We find that for $\epsilon\lesssim 0.011$, where the exact cutoff also depends on the coherence length of the field, the ULDM signal no longer observable due to the large deviations from the mass-radius relation caused by binaries. When the signal is not significant, the (Moran's I)$\times$(Z-Score) curve displays the same qualitative behavior as Fig. \ref{fig:curves_dust_10}.

\subsection{Including Measurement Noise} \label{sec:measurement}

\indent The measured radii, surface gravities, and temperatures in the \citet{Crumpler_2025} catalog have corresponding uncertainties due to limitations in our ability to perfectly measure the physical parameters of DA WDs. The noise from inaccurate measurements can impact our ability to characterize the ULDM signal of Sec. \ref{sec:ULDM}, and so we simulate this effect. The typical distance-dependent uncertainty on radius, surface gravity, and temperature measurements is given by the median of the \texttt{e\_radius\_phot\_full}, \texttt{e\_logg\_prf\_coadd\_full}, and \texttt{e\_teff\_phot\_full} columns of the clean catalog from Sec. \ref{sec:catalog}, respectively, for each $50$ pc distance bin. In the nearest $0$ to $50$ pc distance bin, the median uncertainties are $0.0005~R_\odot$, $0.11$ dex, and $200$ K for radius, surface gravity, and temperature measurements, respectively. In the farthest $950$ to $1000$ pc distance bin, these increase to $0.003~R_\odot$, $0.12$ dex, and $975$ K. We divide these typical uncertainties by an \texttt{improvement\_factor} parameter to investigate how our ability to detect ULDM effects on WD structure may improve with higher quality data or better measurement procedures. Additionally, \citet{Crumpler_2024} finds evidence that these uncertainties might be overestimated, so using the full typical uncertainties (\texttt{improvement\_factor}=1) might be overly pessimistic. 

\indent We repeat the steps of Sec. \ref{sec:init_param}, drawing an initial sample of WD parameters with no sources of noise. We modify the simulated radii and surface gravities according to Sec. \ref{sec:ULDM}. Then, we adjust the observed radius, surface gravity, and temperature of each WD by drawing a random number from a Gaussian distribution with $\mu=0$ and $\sigma=1$, multiplying the random number by the typical uncertainty on that parameter divided by the \texttt{improvement\_factor}, and adding the result to the measured value of the radius, surface gravity, or temperature. We then compute the spatial correlation as in Sec. \ref{sec:moransI} for $1 \leq\Delta x \leq 1000$ pc and $10^{-10} \leq  \epsilon_\text{max} \leq 0.5$. When using \texttt{improvement\_factor}$=10$, the ULDM signal is overcome by measurement noise and is not observable for $\epsilon_\text{max}\lesssim 0.034$. Decreasing the \texttt{improvement\_factor} parameter means increasing the measurement errors and causes the ULDM signal to disappear at larger $\epsilon$. For \texttt{improvement\_factor}=1, when the uncertainties are similar to the nominal stated uncertainties in the clean catalog, the ULDM signal is only observable for $\epsilon_\text{max}\gtrsim 0.23$, where the exact cutoff depends on the coherence length of the field.  When the signal is not significantly detectable, the (Moran's I)$\times$(Z-Score) curve displays the same qualitative behavior as Fig. \ref{fig:curves_dust_10}.

\subsection{Including Distance Uncertainty} \label{sec:distance}

\indent The \citet{BailerJones_2021} distance to each WD has a corresponding uncertainty which can impact our ability to characterize the ULDM signal of Sec. \ref{sec:ULDM}, and so we simulate this effect. The typical distance-dependent uncertainty on the distance is given by the median of the difference between the \texttt{r\_hi\_geo} and \texttt{r\_lo\_geo} columns of the clean catalog from Sec. \ref{sec:catalog}, for each $50$ pc distance bin. In the nearest $0$ to $50$ pc distance bin the median uncertainty is $0.18$ pc, and in the farthest $950$ to $1000$ pc distance bin this increases to $320$ pc. Because distances are used in measuring the photometric radii and temperatures of WDs, this distance uncertainty can impart additional uncertainty in these measurements. However, \citet{Crumpler_2025} already take this into account when characterizing the full error on their photometric radii and temperatures, and so this effect is already accounted for in Sec. \ref{sec:measurement}.

\indent We repeat the steps of Sec. \ref{sec:init_param}, drawing an initial sample of WD parameters with no sources of noise. We modify the simulated radii and surface gravities to account for ULDM according to Sec. \ref{sec:ULDM}. Then, we adjust the observed distance of each WD by drawing a random number from a Gaussian distribution with $\mu=0$ and $\sigma=1$, multiplying the random number by the typical distance uncertainty, and adding the result to the initial value of the distance. We then compute the spatial correlation as in Sec. \ref{sec:moransI} for $1 \leq\Delta x \leq 1000$ pc and $10^{-10} \leq  \epsilon_\text{max} \leq 0.5$. We find that distance uncertainty does not have a strong impact on our ability to detect ULDM, and the ULDM signal is visible for all simulated maximum deviations of the observed radius ($\epsilon_\text{max} \geq10^{-10}$). This is because the overall median distance uncertainty is $21$ pc, and a distance uncertainty of this size is smaller than most coherence lengths considered here. So, distance uncertainty does not often result in WDs being mistakenly identified with an incorrect ULDM field interference patch.

\begin{figure}[h!]
\begin{center}
\includegraphics[scale=0.25]{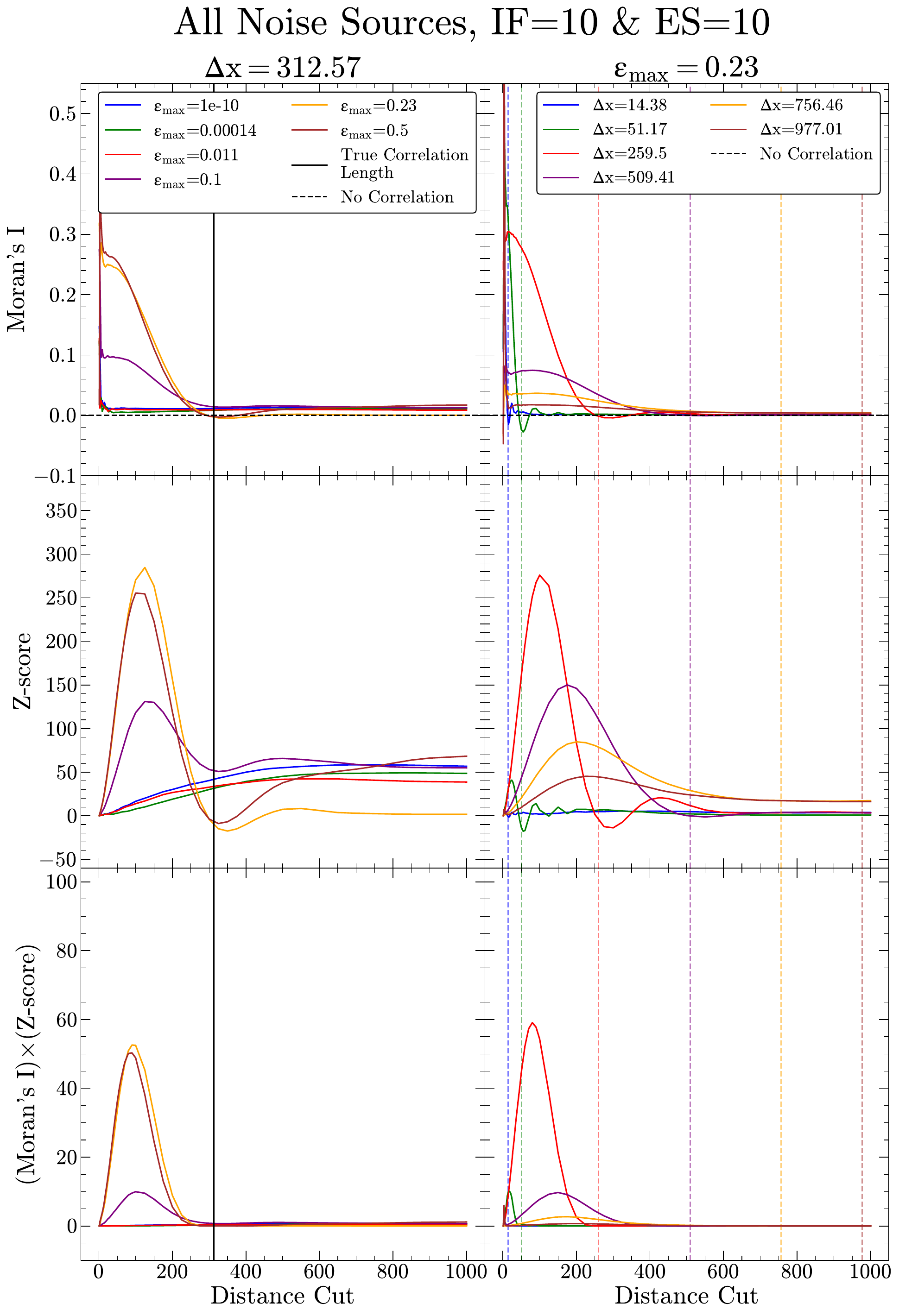}
\caption{Same as Figs. \ref{fig:curves_no_noise} and \ref{fig:curves_dust_10}, but with noise due to thin hydrogen envelope contamination, dust, binary contamination, measurement noise, and distance uncertainty added. We improve measurement uncertainties and reduce the strength of the effect of dust by a factor of 10. For sufficiently small $\epsilon_\text{max}$, the ULDM signal is washed out by the effect of various noise sources, resulting in a monotonically increasing of the (Moran's I)$\times$(Z-score) curve. \label{fig:curves_all_noise_IF10_ES10}}
\end{center}
\end{figure}

\subsection{Simulation Output For Multiple Noise Sources} \label{sec:noisy}

\indent We combine the noise effects of Secs. \ref{sec:thinH} - \ref{sec:distance} to produce a more realistic simulation of various effects impacting the detectability of the spatial correlation in WD deviations from the mass-radius relation imparted by ULDM. Fig. \ref{fig:curves_all_noise_IF10_ES10} shows the outputs of some of these simulations including the effects of ULDM, thin hydrogen envelope contamination, dust, binary contamination, measurement noise, and distance uncertainty where we  improve measurement uncertainties and reduce the strength of the effect of dust by an order of magnitude (\texttt{improvement\_factor}=10, \texttt{effect\_strength}=10). For $\epsilon_\text{max}\lesssim 0.034$, the ULDM signal is overcome by other noise sources and is no longer observable. Decreasing the \texttt{improvement\_factor} or \texttt{effect\_strength} parameters causes the ULDM signal to disappear at larger $\epsilon_\text{max}$. For \texttt{improvement\_factor}=1 and \texttt{effect\_strength}=1, for \texttt{improvement\_factor}=1 and \texttt{effect\_strength}=10, and for \texttt{improvement\_factor}=10 and \texttt{effect\_strength}=1, the ULDM signal is only observable for $\epsilon_\text{max}\gtrsim 0.37$, $\epsilon_\text{max}\gtrsim 0.5$, and $\epsilon_\text{max}\gtrsim 0.23$, respectively, where the exact cutoff depends on the coherence length of the field. Thus, the effects of dust and measurement noise have the strongest impacts on the observability of the signal.

\begin{figure}[h!]
\begin{center}
\includegraphics[scale=0.25]{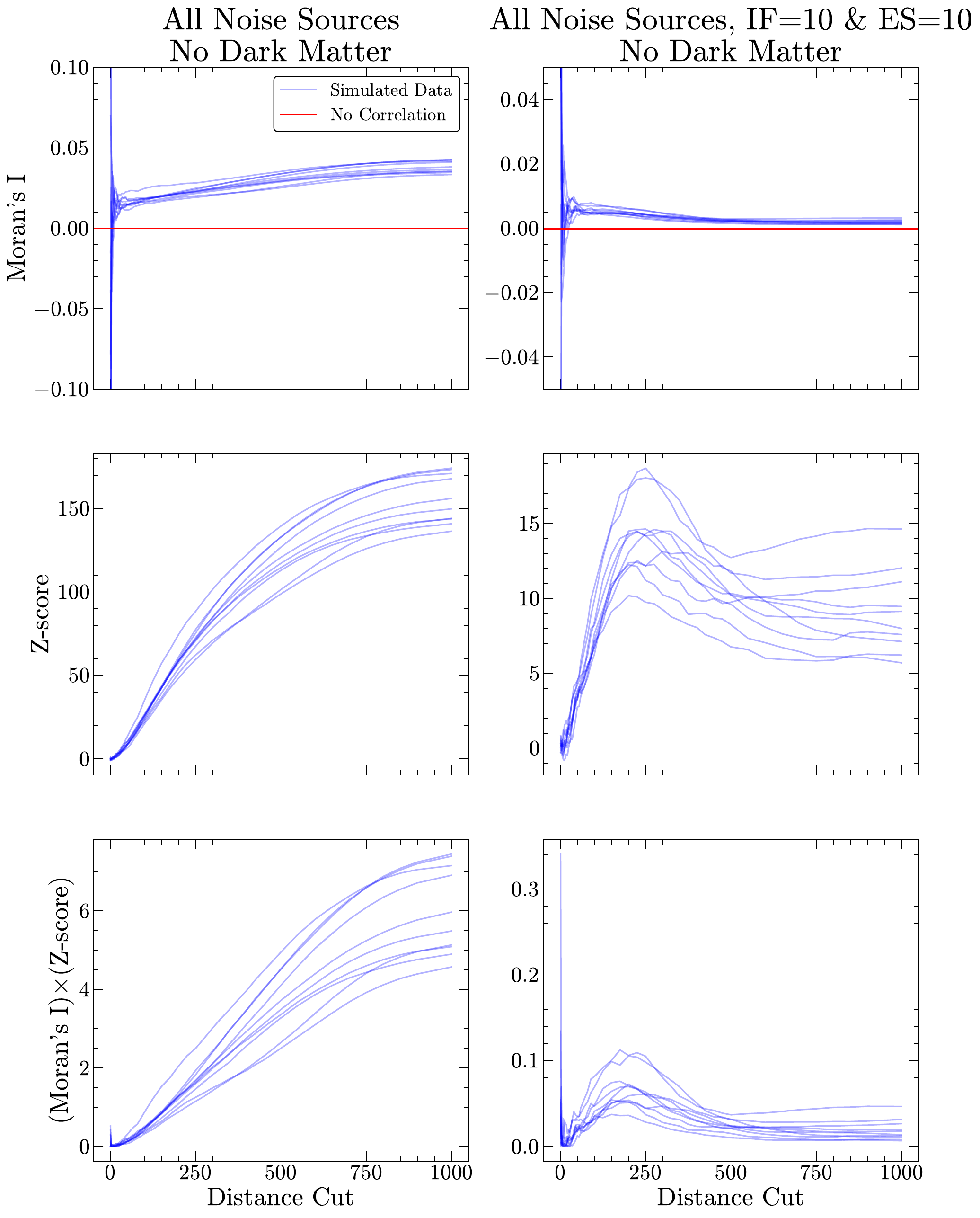}
\caption{The spatial autocorrelation signal from a simulated catalog of 10,207 WDs with no ULDM effects. The left column shows the Moran's I and Z-score statistic when all noise sources are included, and the right column shows the same but has \texttt{improvement\_factor}=\texttt{effect\_strength}=10. We find that for some configurations of \texttt{improvement\_factor} and \texttt{effect\_strength}, we can produce an ULDM-like signal even when not including ULDM, but this signal is much weaker than in simulations including ULDM. \label{fig:noULDM}}
\end{center}
\end{figure}

\subsection{Simulation Output For Multiple Noise Sources with No ULDM} \label{sec:noisy_no_uldm}

\indent We repeat the simulations of Sec. \ref{sec:noisy} without including ULDM to investigate whether ULDM-like signals can be created by other sources. We run these simulations with all noise sources (thin hydrogen envelope contamination, dust, binary contamination, measurement noise, and distance uncertainty) and with all noise sources and decreasing both measurement noise and the strength of dust effects by an order of magnitude (\texttt{improvement\_factor}=\texttt{effect\_strength}=10).

\indent The outputs of 10 realizations of each of these simulations are displayed in Fig. \ref{fig:noULDM}. For the simulations with all noise sources there is no ULDM-like signal. Rather, the presence of dust creates a positive spatial correlation at all separations which results in a monotonically increasing Z-score and (Moran's I)$\times$(Z-score) curve. For the simulations with  all noise sources and \texttt{improvement\_factor}=\texttt{effect\_strength}=10, we find that the combination of dust and other noise sources produces a peak in the (Moran's I)$\times$(Z-score) at $\sim 250$ pc that appears similar to an UDLM signal. This peak results from measurement noise washing out the positive correlation due to dust at large WD separations. This peak value is $\lesssim 0.1$, which is much lower than the peaks created by ULDM which can be as high as 200. This ULDM-like signal is highly sensitive to the balance between dust effects and measurement noise, and thus is also sensitive to the choice of \texttt{improvement\_factor} and \texttt{effect\_strength}.

\section{Reconstructing the Field coherence length} \label{sec:recon_corr_len}

\indent In the figures of Sec. \ref{sec:simulation}, it is evident that the shape of the (Moran's I)$\times$(Z-score) curve corresponds to the coherence length of the ULDM background field, if the variations in the observed radius are sufficiently large for the signal to be observable. The peak of the (Moran's I)$\times$(Z-score) function typically occurs at spatial separations that are $1/3-1/2$ of the coherence length, and the function returns to a background level right at or at slightly shorter distances than the true coherence length. Thus, even though there is a strong relationship between these features and the true coherence length, there is not a one-to-one relationship that can be easily described analytically. This situation is well-suited for employing a deep learning approach to create a regression between the shape of the curve and the true coherence length.

\indent For each of the noise configurations of Sec. \ref{sec:simulation}, we use the one-dimensional (Moran's I)$\times$(Z-score) curve labeled with the true coherence length of each simulation to train a convolutional neural network (CNN) to reconstruct the ULDM field coherence length given the shape of the input curve for that noise configuration. Typically applied to two-dimensional data, CNNs are a type of deep learning framework specialized for feature extraction to detect things like shapes, edges, and textures on image-like data. The one-dimensional adaptation of the CNN framework is well-suited for ordered data to extract patterns from the input, such as how the shape of a curve corresponds to a particular coherence length.

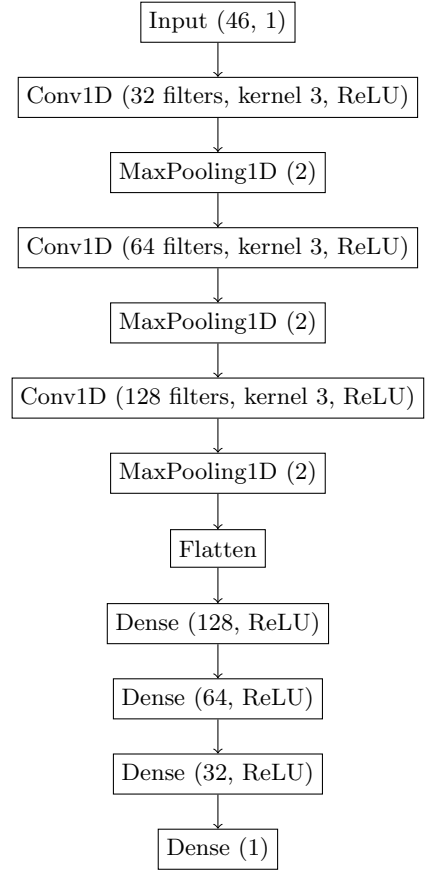
\begin{figure}[ht!]
    \centering
    \begin{tikzpicture}[node distance=1cm, scale=0.8] 
        \node (input) [draw, rectangle, minimum height=0.5cm, minimum width=1cm] {Input (46, 1)};
        
        \node (conv1) [draw, rectangle, below of=input, minimum height=0.5cm, minimum width=1cm] {Conv1D (32 filters, kernel 3, ReLU)};
        
        \node (pool1) [draw, rectangle, below of=conv1, minimum height=0.5cm, minimum width=1cm] {MaxPooling1D (2)};
        
        \node (conv2) [draw, rectangle, below of=pool1, minimum height=0.5cm, minimum width=1cm] {Conv1D (64 filters, kernel 3, ReLU)};
        
        \node (pool2) [draw, rectangle, below of=conv2, minimum height=0.5cm, minimum width=1cm] {MaxPooling1D (2)};
        
        \node (conv3) [draw, rectangle, below of=pool2, minimum height=0.5cm, minimum width=1cm] {Conv1D (128 filters, kernel 3, ReLU)};
        
        \node (pool3) [draw, rectangle, below of=conv3, minimum height=0.5cm, minimum width=1cm] {MaxPooling1D (2)};
        
        \node (flatten) [draw, rectangle, below of=pool3, minimum height=0.5cm, minimum width=1cm] {Flatten};
        
        \node (dense1) [draw, rectangle, below of=flatten, minimum height=0.5cm, minimum width=1cm] {Dense (128, ReLU)};
        
        \node (dense2) [draw, rectangle, below of=dense1, minimum height=0.5cm, minimum width=1cm] {Dense (64, ReLU)};
        
        \node (dense3) [draw, rectangle, below of=dense2, minimum height=0.5cm, minimum width=1cm] {Dense (32, ReLU)};
        
        \node (output) [draw, rectangle, below of=dense3, minimum height=0.5cm, minimum width=1cm] {Dense (1)};
        
        \draw[->] (input) -- (conv1);
        \draw[->] (conv1) -- (pool1);
        \draw[->] (pool1) -- (conv2);
        \draw[->] (conv2) -- (pool2);
        \draw[->] (pool2) -- (conv3);
        \draw[->] (conv3) -- (pool3);
        \draw[->] (pool3) -- (flatten);
        \draw[->] (flatten) -- (dense1);
        \draw[->] (dense1) -- (dense2);
        \draw[->] (dense2) -- (dense3);
        \draw[->] (dense3) -- (output);
    \end{tikzpicture}
    \caption{CNN Architecture for best-performing model trained for 400 epochs with Adam optimizer, mean squared error loss function, and a batch size of 32. The input is the one-dimensional (Moran's I)$\times$(Z-score) curve, which is a series of 46 ordered points, with each point corresponding to a particular distance cutoff. The output of the final dense layer is the measured coherence length in pc.}\label{fig:cnn_config}
\end{figure}

\indent For each of the simulation configurations of Sec. \ref{sec:simulation}, we create a training data set of 2,921 simulations and a testing data set of 256 simulations with $1 \leq\Delta x \leq 1000$ pc and $10^{-10} \leq  \epsilon_\text{max} \leq 0.5$. For the training data set we simulate (Moran's I)$\times$(Z-score) across the full range of $\epsilon_\text{max}$ while for the testing data set we focus on larger $\epsilon_\text{max}$ since we find that for most simulations the ULDM effect is no longer observable for $\epsilon_\text{max}<0.034$ in Sec. \ref{sec:simulation}. For each simulation configuration, we build a variety of CNN models with \texttt{tensorflow.keras.models.Sequential}. We vary the number of layers in the model, the number of training epochs, the maximum pooling parameter which reduces the dimensionality of the feature maps in pooling layers, the kernel size parameter which defines the convolution window size in convolution layers, and the training batch size. For each simulation and CNN configuration, we train the model on the 2,921 (Moran's I)$\times$(Z-score) curves, normalized by the maximum of the absolute value of the curve and labeled with the true coherence length. We then apply the trained model to the testing data set of (Moran's I)$\times$(Z-score) curves to produce a detectability heatmap. This heatmap shows the fractional error between the true coherence length of the ULDM field and the coherence length measured by the CNN as a function of the true coherence length and the maximum variation of the observed radius. We determine the optimal CNN configuration by identifying the heatmap which minimizes the number of testing simulations with fractional error $>10\%$ and by investigating the median absolute error between the true and predicted coherence lengths for each testing simulation. For most simulation configurations, the best performing CNN architecture is that described in Fig. \ref{fig:cnn_config}. This architecture performs well even when applied to simulations for which it is not the best performing architecture, so, for consistency, we choose to employ this architecture for all final CNN models for all simulations.

\begin{figure*}[h!]
\begin{center}
\includegraphics[scale=0.3]{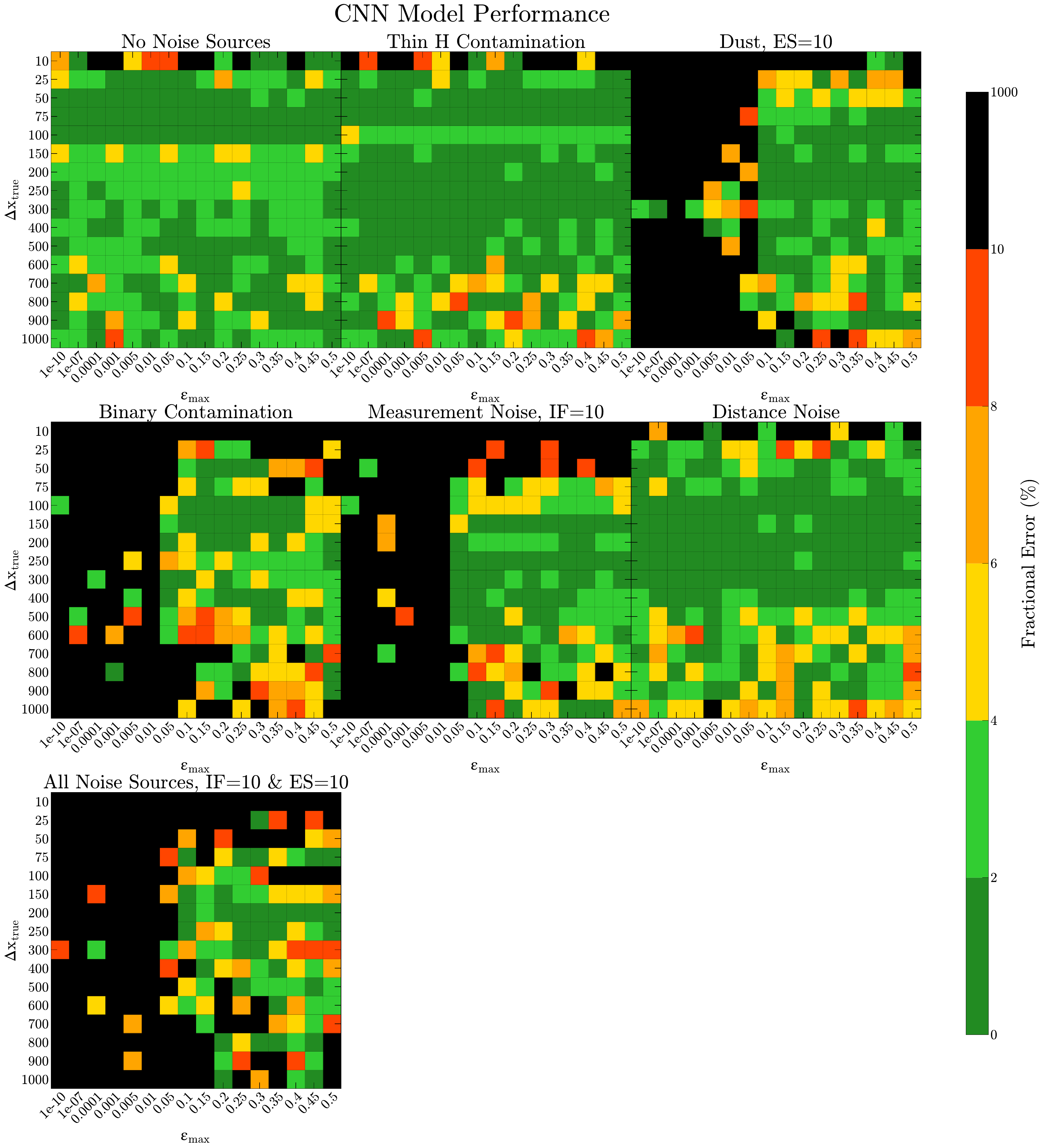}
\caption{Heatmaps for each simulation configuration showing the fractional error between the true coherence length of the ULDM field and the coherence length measured by the CNN as a function of the true coherence length and the maximum variation of the observed radius. Measurement noise and dust have the greatest impact on the observability of the ULDM signal. Generally, the signal is harder to reconstruct for small $\epsilon_\text{max}$ and for very small or very large coherence lengths ($\Delta x_\text{true}\sim 10$ or $1000$ pc). The presence of dust creates a preference for measured correlation lengths of $\sim300$ pc. \label{fig:cnn}}
\end{center}
\end{figure*}

\indent In Fig. \ref{fig:cnn}, we show these detectability heatmaps for each of the simulation configurations of Sec. \ref{sec:simulation}. All heatmaps in Fig. \ref{fig:cnn} use the CNN architecture of Fig. \ref{fig:cnn_config}. For the heatmaps corresponding to simulations with no noise sources, only thin hydrogen envelope contamination, and only distance noise, we find that the ULDM signal is detectable across the full range of $10^{-10} \leq  \epsilon_\text{max} \leq 0.5$ and $10\leq\Delta x\leq 1000$. For coherence lengths of $10<\Delta x< 1000$, the CNN is able to accurately reconstruct the ULDM field coherence length, with most measured coherence lengths being within $4\%$ of the true value. At a coherence length of $\Delta  x=10$ or $1000$, the fractional error is higher and CNN has more difficulty accurately determining the coherence length, with some cases in which the CNN fails to accurately measure the coherence length (fractional error $>10\%$).

\indent For the heatmaps corresponding to simulations with only dust effects, only binary contamination, only measurement noise, and all noise sources, we find that for small $\epsilon_\text{max}$ and some coherence lengths, the ULDM coherence length is no longer able to be accurately measured by the CNN. For dust only, the CNN fails for $\epsilon_\text{max}<0.1$ and for $\Delta x< 25$ pc. For binary contamination only, the CNN fails for $\epsilon_\text{max}<0.05$ and for $\Delta x<50$ pc or $\Delta x>900$ pc. For measurement noise only, the CNN fails for $\epsilon_\text{max}<0.05$ and for $\Delta x<75$ pc. For all noise sources, the CNN fails for $\epsilon_\text{max}<0.1$ and for $\Delta x<75$ pc or $\Delta x>800$ pc. In heatmaps with dust effects, there is a preference for coherence lengths of $\Delta x\sim300$ pc. From Fig. \ref{fig:noULDM}, we see that this preference arises from the ULDM-like signal created by dust at these coherence lengths. Overall, across all simulations the impact of ULDM on WD structure can be measured for a wide range of ULDM field coherence lengths and maximum observed radius variations with these methods. When all noise sources are turned on, relatively large variations in the radius ($\epsilon_\text{max}\gtrsim0.1$) are needed in order for the coherence length to be reconstructed by the CNN. 

\section{Searching for ULDM in Real Data} \label{sec:data}

\begin{figure}[h!]
\begin{center}
\includegraphics[scale=0.4]{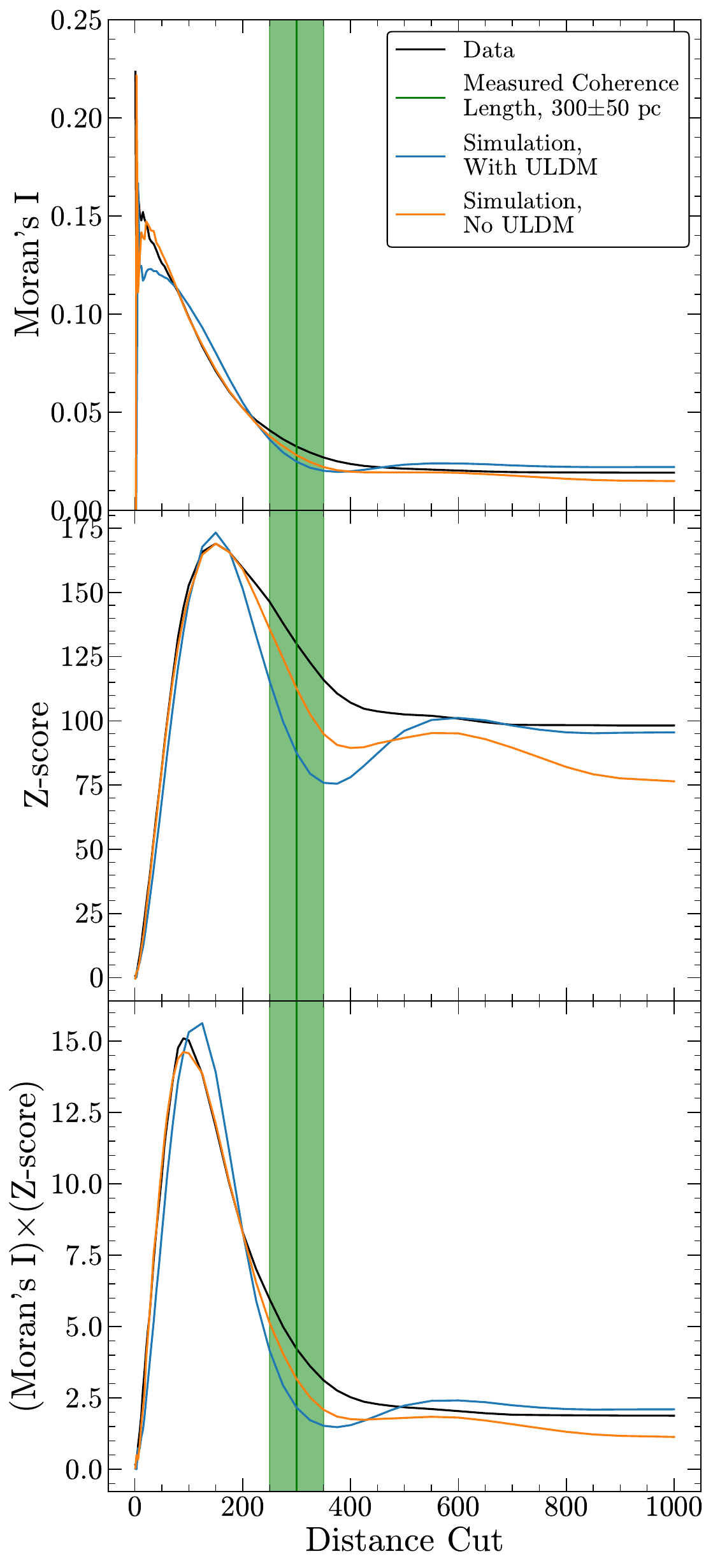}
\caption{The spatial autocorrelation signal from the real data of the clean catalog of Sec. \ref{sec:catalog} (black), from a simulation with ULDM (blue), and from a simulation with purely distance-dependent biases and offsets and no ULDM (orange). The green line and shaded region show the measured coherence length and uncertainty on that coherence length, respectively, for the black curve. Both the ULDM and the non-ULDM models reproduce the data well. Thus, there is a degeneracy between spatial correlations due to distance effects and to ULDM.\label{fig:curves_real_data}}
\end{center}
\end{figure}

\indent Using the methods of Sec. \ref{sec:moransI}, we calculate the Moran's I statistic and Z-score of the statistic for correlations in the deviations from the WD mass-radius relation as a function of WD separation for the 10,207 unique DA WDs contained in the clean catalog of Sec. \ref{sec:catalog}. The resulting Moran's I and Z-score curves as a function of distance cut are displayed in Fig. \ref{fig:curves_real_data}. The curves show a clear positive spatial correlation among WDs with separations $\lesssim500$ pc.

\indent To check the robustness of this signal, we re-measure the spatial correlation for the same catalog without extinction corrections. The process for creating this uncorrected catalog is described in Sec. \ref{sec:dust}. Excluding extinction corrections reduces the height of the peak in the (Moran's I)$\times$(Z-score) curve slightly, from $\sim 15$ to $\sim12$, and removes the tail at large spatial separations, but does not change the presence or shape of the signal. Additionally, we check for sampling bias by randomly sampling half of the extinction-corrected catalog without replacement and re-measuring the spatial correlation. We find that the the height of the peak in the (Moran's I)$\times$(Z-score) is reduced to $\sim7$ since fewer WDs are included in the measurement, but the shape of the curve is unchanged. Finally, we check for geometric bias by taking the real positions of the WDs in the catalog and replacing the measured deviation from the mass-radius relation with Gaussian noise. We use a Gaussian distribution with a standard deviation set by the standard deviation of the real deviations from the clean catalog, and test using a mean of zero and a mean equal to the mean deviation of the clean catalog. In both cases, this washes out the signal, resulting in a monotonically increasing (Moran's I)$\times$(Z-score) curve. Thus, we conclude that the spatial correlation in Fig. \ref{fig:curves_real_data} shows the presence of a real positive spatial correlation among the WDs contained in the clean catalog. 

\indent We apply the CNN model with all noise sources and \texttt{improvement\_factor}=\texttt{effect\_strength}=10, shown in Fig. \ref{fig:cnn}, to the (Moran's I)$\times$(Z-Score) curve in Fig. \ref{fig:curves_real_data} to measure an UDLM field coherence length. In order to marginalize over the variation in the measured coherence length induced by random differences in training the model and by the choice of training data, we bootstrap the training data set and re-train 300 versions of the CNN. We apply all 300 versions of the CNN to the (Moran's I $\times$ Z-Score) curve from the clean catalog, and calculate the mean and standard deviation of the measured coherence length. We measure a coherence length of $300\pm50$ pc, which, if the autocorrelation peak is due to the presence of ULDM, corresponds to an ULDM mass of $m_\text{DM}\sim 2\times10^{-23}$ eV.

\indent We run $\sim15,000$ simulations with \texttt{improvement\_factor} $\in(5,15)$, \texttt{effect\_strength} $\in(5,15)$, $\Delta x\in(250,350)$ pc, and $\epsilon_\text{max}\in(0.01,0.5)$, and find the simulation that best matches the (Moran's I $\times$ Z-Score) curve from the clean catalog. To find this best-matching simulation, we identify the simulation for which the maximum absolute difference between the simulation and the real data (Moran's I)$\times$(Z-Score) curves is minimized. The best-matching simulation has \texttt{improvement\_factor}=10, \texttt{effect\_strength}=5, $\Delta x$= 345 pc, and $\epsilon_\text{max}$=0.14, and we plot the autocorrelation function from this simulation in Fig. \ref{fig:curves_real_data}. This best-matching simulated curve reconstructs the location and amplitude of the real data well, given the high peak of $\sim 15$ in the (Moran's I $\times$ Z-Score) curve of the real data. However, the shape of the simulated ULDM (Moran's I)$\times$(Z-score) curve is narrower than the real data. This narrower signal might be caused by the simple model for the ULDM background, in which every interference granule is the exact same size. In a more realistic simulation, the typical size of the interference granules would be set by the coherence length, but there would be a broader distribution of granule sizes about this typical value, which would likewise broaden the ULDM signal.

\indent In Fig. \ref{fig:noULDM}, we find that our simulations can produce an ULDM-like signal in the absence of ULDM due to the balance between offsets from dust and noise from measurement uncertainties. However, this signal is two orders of magnitude too weak compared to the  (Moran's I)$\times$(Z-Score) curve from the clean catalog shown in Fig. \ref{fig:curves_real_data}. In the simulations of Sec. \ref{sec:simulation} and in Fig. \ref{fig:noULDM}, we include distance-dependent biases and uncertainties in construction of the simulated WD sample. However, these distance-dependent effects may not be fully accounted for. We investigate this further by creating a simulation with no ULDM and purely distance-dependent effects on the deviations from the mass-radius relation.

\indent In this distance-dependent simulation with no ULDM, we first repeat the steps of Sec. \ref{sec:init_param}, drawing an initial sample of WD parameters with no sources of noise. We then calculate the initial deviations from the mass-radius relation according to Sec. \ref{sec:moransI}. We modify these initial deviations with a distance-dependent offset and noise level. The offset determines if measured WD radii at that distance are systematically too large or too small relative to the theoretical mass-radius relation and the noise level indicates the uncertainty on those measurements. For each simulated WD, we draw a distance cutoff from a Gaussian distribution with a mean and standard deviation taken as inputs to the simulation. We also input another, larger distance cutoff in the simulation. For each WD, if the distance is nearer than the drawn distance cutoff parameter, then the noise and offset are set to one value, and if the distance is farther than that cutoff, then the noise and offset are set to a different value. For the offsets, we also include a third value for WDs with very large distances, beyond the second distance cutoff. Each of these values are taken as inputs to the simulation. Given the noise and offset for that WD, we draw a random deviation about the initial deviation from a Gaussian distribution with a mean that is the sum of the initial deviation and the offset parameter and a standard deviation that is the noise parameter. We then calculate the spatial correlation statistics for the simulation according to Sec. \ref{sec:moransI}. 

\indent We run $\sim$ 9,000 of these distance-dependent simulations with no ULDM for a mean distance cutoff $\in(120,160)$ pc, a standard deviation of the distance cutoff $\in(5,40)$ pc, a far distance cutoff $\in(400,600)$ pc, a nearby noise level $\in(1,10)$ \%, a far noise level $\in(1,30)$ \%, a nearby offset $\in(50,90)$ \%, a medium distance offset $\in(0,20)$ \%, and a far offset $\in(50,90)$ \%. We find the simulation that best matches the (Moran's I)$\times$(Z-Score) curve from the clean catalog. The best-matching simulation has a mean distance cutoff of 120 pc, a standard deviation of the distance cutoff of 20 pc, a far distance cutoff of 400 pc, a nearby noise level of 5\%, a far noise level of 10\%, a nearby offset of 70\%, a medium distance offset of 0\%, and a far offset of 70\%, and we plot the results of this simulation in Fig. \ref{fig:curves_real_data}. Generally, we find that the observational signal of Fig. \ref{fig:curves_real_data} can be recreated by a non-ULDM simulation with large offsets nearby ($d\lesssim150$ pc) and very far away ($d\gtrsim400$ pc) as well as low uncertainties nearby and higher uncertainties far away. The low uncertainty nearby and high uncertainty far away as well as the high offset very far away are in line with our expectations, since distant things are harder to measure. The large offset nearby is indicative of observation bias because certain classes of WDs are more likely to be observed nearby. For example, cool and massive WDs are harder to observe and thus are more likely to be located at shorter distances.

\begin{figure}[h!]
\begin{center}
\includegraphics[scale=0.25]{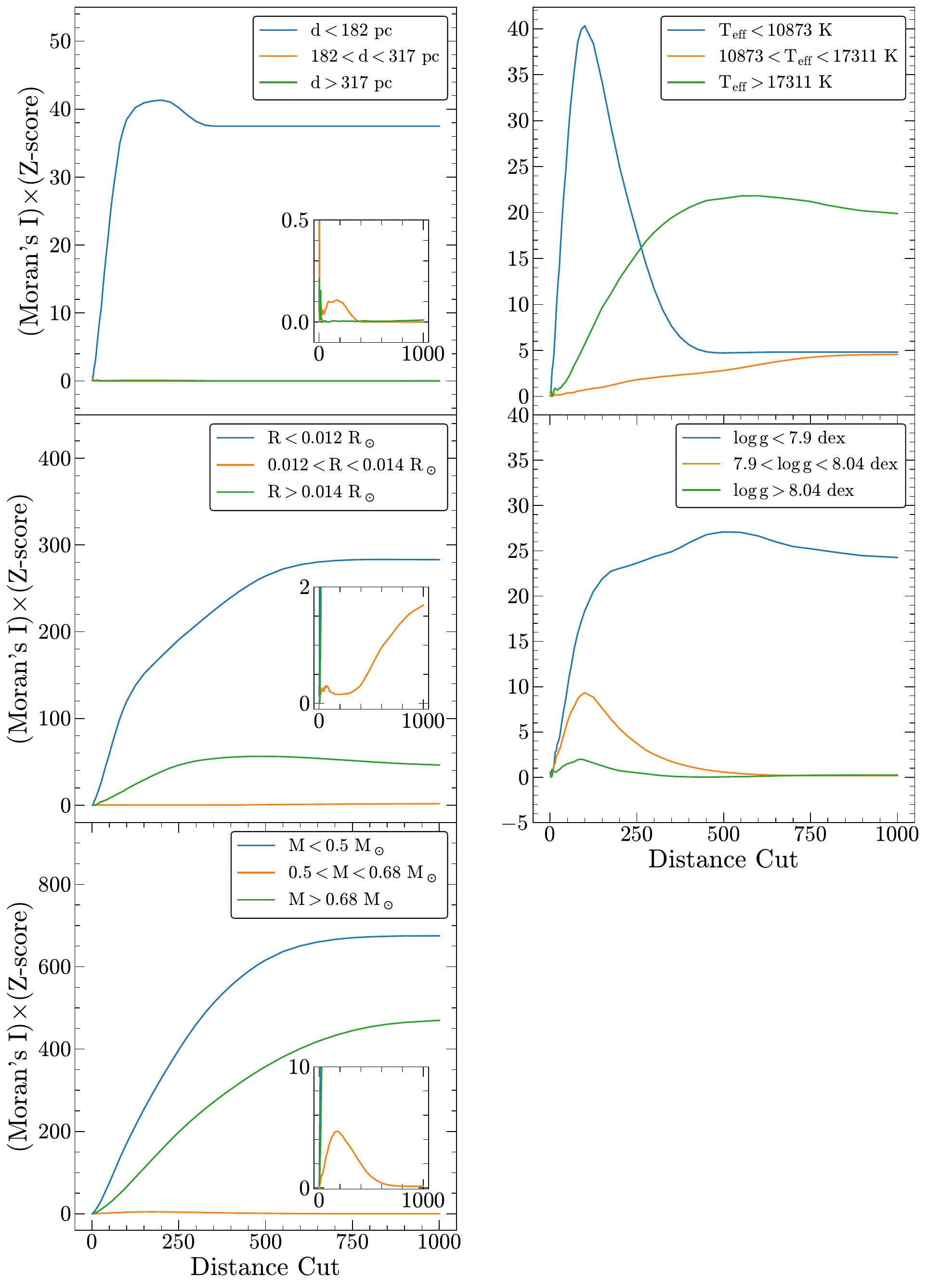}
\caption{From left to right and top to bottom, the spatial autocorrelation signal from the real data of the clean catalog of Sec. \ref{sec:catalog} as a function of distance, effective temperature, radius, surface gravity, and mass bin. For each parameter, the clean catalog is divided into three equal-sized sub-catalogs of low (blue), medium (orange), and high (green) values of that parameter. Based on the peaked structures in the sub-catalogs binned by distance and by effective temperature, we conclude that the positive correlation in Fig. \ref{fig:curves_real_data} is due to observational bias. \label{fig:curves_bin_params}}
\end{center}
\end{figure}

\indent Both the simulations with ULDM and without ULDM, but with distance-dependent effects reproduce the (Moran's I)$\times$(Z-Score) curve from the clean catalog well. Thus, there is a degeneracy between distance effects and ULDM necessitating careful consideration of distance-dependent observational biases. To pursue this further, we investigate how different distance, effective temperature, radius, surface gravity, and mass cuts impact the measured spatial correlation from the clean catalog. For each parameter, we divide the catalog into three equal-sized sub-catalogs corresponding to low, medium, and high values of that parameter. Each sub-catalog contains $\sim 3,400$ WDs. We then compute the spatial correlation for each of these sub-catalogs for each parameter, and these (Moran's I)$\times$(Z-Score) curves are shown in Fig. \ref{fig:curves_bin_params}.

\indent Fig. \ref{fig:curves_bin_params} shows that the observed spatial correlation in Fig. \ref{fig:curves_real_data} is due to observational bias. When dividing the catalog based on distance, we find that the positive spatial correlation only occurs for nearby WDs ($d<182$ pc), and that the signal is almost washed out for WDs at intermediate distances and completely washed out for the most distant WDs ($d>317$ pc). Dividing the catalog based on effective temperature shows why this is the case. Fig. \ref{fig:curves_bin_params} shows that the peaked structure in the positive spatial correlation occurs very strongly for low effective temperature ($T_\text{eff}<10873$ K) WDs, while for intermediate effective temperature WDs there is a slight peaked structure and for high effective temperature WDs any signal is washed out. Looking at dividing the catalog based on radius, surface gravity or mass, we see no peaked structure as strong as the one from dividing based on effective temperature. Thus, the peaked structure in Fig. \ref{fig:curves_real_data} is being driven by low temperature WDs. Intuitively, this makes sense, as low effective temperature WDs are dimmer and thus mostly observed nearby and these WDs are more poorly understood than high effective temperature WDs, leading to offsets from theoretical expectations \citep{ElBadry_2018, Bergeron_2019}. Thus, the positive spatial correlation among the low temperature WD curve can be explained by an offset between measurements and theory, but mostly for nearby WDs, and the presence of few low temperature WDs with noisy measurements at further distances causing the decrease in the (Moran's I)$\times$(Z-score) curve at distances greater than $\sim$500 pc. 

\indent The positive correlation due to nearby, low temperature WDs in the absence of ULDM does not appear in our initial simulations of Sec. \ref{sec:simulation}, even though those simulations contain distance-dependent effects. This is because those simulations assume that theory and observation agree to within the stated measurement uncertainties. The positive correlation in Fig. \ref{fig:curves_bin_params} shows that assumption is not true, and that the measured masses and radii for low temperature WDs in the \citet{Crumpler_2025} catalog do not agree with the theoretical mass-radius relation. 

\indent In the curves of Fig. \ref{fig:curves_bin_params} divided by radius, we see no evidence of a peaked structure and all of the curves are dominated by noise. For the curves divided by surface gravity, the curve for the lowest surface gravity WDs is dominated by noise while the other two curves show peaked structures. The strongest peak results from WDs in the middle surface gravity range, between 7.90 and 8.04 dex. In \citet{Crumpler_2025}, the authors acknowledge that of all their measurement procedures, their spectroscopic surface gravity routine has the most room for improvement. Their random forest routine may bias WD surface gravities towards the center of the WD mass distribution, where there is more training data available. This would result in an offset from theoretical expectations near the peak of the WD surface gravity distribution, which occurs in the middle surface gravity range and explains the structure in Fig. \ref{fig:curves_bin_params}. The curves divided by mass are dominated by noise except for the intermediate mass range, between 0.50 and 0.68 $M_\odot$, which shows a slight peaked structure due to the corresponding peak in intermediate range surface gravities.

\section{Discussion and Conclusions} \label{sec:conc}

\indent In this paper, we have characterized the spatial correlation in deviations of WDs from their well-known mass-radius relation imparted by an ULDM background field. Motivated by how an ULDM-induced variation in the electron mass would change the radius of a WD, we consider the effects of generic ULDM-induced spatial variations in WD structure of the form $R(\epsilon)=(1+\epsilon)R_0$, where $\epsilon$ encompasses the combination of the amplitude of the ULDM scalar field variation and the strength of the coupling between the ULDM field and the Standard Model. If two WDs are within the same ULDM coherence patch, the radii of these stars are changed in a correlated way. Given that observed DA WDs are typically separated by distances of 10 - 1000 pc, WDs can probe ULDM field coherence lengths on this scale, which corresponds to probing dark matter masses of $m_\text{DM}\sim 6\times10^{-24}-6\times10^{-22}$ eV.

\indent Using the \citet{Crumpler_2025} catalog, we create a sample of 10,207 unique DA WDs with high-quality spectroscopic and photometric measurements within 1000 pc. This clean catalog serves as the input for our Monte Carlo simulation, which we use to model the detectability of the ULDM spatial correlation signal given effects from thin hydrogen envelope contamination, dust, binary contamination, measurement noise, and distance uncertainty. We find that the best indicator of the underlying coherence length of the ULDM field is the product of the Moran's I statistic, which characterizes the extent of the positive correlation, with the Z-score of the statistic, which indicates the statistical significance of the positive correlation. When the ULDM signal is not washed out by noise, the (Moran's I)$\times$(Z-score) curve has a characteristic peaked shape with a maximum at $\sim 1/2-1/3$ the coherence length of the field and decreases to a plateau at around the coherence length. When the ULDM signal is washed out by noise, the (Moran's I)$\times$(Z-score) curve monotonically increases. 

\indent We create $\sim 3,000$ simulations of the ULDM signal for each of a variety of noise configurations, and use these simulations as training data sets to build CNNs capable of measuring the background coherence length of an ULDM field given a (Moran's I)$\times$(Z-score) curve as a input. We find that, when all noise sources are turned on, relatively large variations in the radius ($\epsilon_\text{max}\gtrsim0.1$) are needed in order for the coherence length to be reconstructed by the CNN. For smaller variations in the radius, the ULDM signal is washed out by noise.

\indent We apply our spatial correlation measurement routine to the clean catalog of 10,207 DA WDs from \citet{Crumpler_2025}, and detect a positive spatial correlation among WDs at moderate separations. This spatial correlation is robust against sampling bias, geometric bias, and the application of extinction corrections. We apply our CNNs trained on simulations  to the measured (Moran's I)$\times$(Z-score) curve from the real data to measure a coherence length of $300\pm50$ pc. We run more simulations with ULDM, and find that the data is best-matched by an ULDM simulation with a coherence length of $\Delta x=345$ pc and a maximum radius variation of $\epsilon_\text{max}=0.14$. This is a large value for the maximum radius variation. 

\indent We find that, even in the absence of ULDM, the combination of an offset from dust, which is a systematic difference between measured WD radii and expected WD radii from the theoretical mass-radius relation, and noise from measurement uncertainties can create an ULDM-like signal in our simulations. However, the amplitude of this signal is two orders of magnitude weaker than what we observe in the real data. We investigate this further by creating a second simulation of purely distance-dependent offsets  and noise in measured WD deviations from the mass-radius relation, which does not include ULDM. We find that the data is also best-matched by a distance-dependent simulation with a mean distance cutoff of 120 pc, a standard deviation of the distance cutoff of 20 pc, a far distance cutoff of 400 pc, a nearby noise level of 5\%, a far noise level of 10\%, a nearby offset of 70\%, a medium distance offset of 0\%, and a far offset of 70\%. As in the best-matching ULDM simulation, these are relatively large offsets between the observed and theoretically expected simulated WD radii. Thus, there is a degeneracy between WD spatial correlations due to distance effects and to ULDM.

\indent In order to break this degeneracy, for each of the distance, effective temperature, radius, surface gravity, and mass parameters, we divide the catalog into three equal-sized sub-catalogs corresponding to low, medium, and high values of that parameter. We then re-measure the spatial correlation for each of these sub-catalogs. Based on the peaked structures in the sub-catalogs binned by distance and by effective temperature, we conclude that the positive correlation in Fig. \ref{fig:curves_real_data} is due to observational bias. In particular, the signal in Fig. \ref{fig:curves_real_data} is driven by low-temperature WDs, and the peaked shape of the low temperature WD spatial correlation curve can be explained by an offset between measurements and theory for nearby cool WDs and the presence of few low temperature WDs with noisy measurements at further distances. Currently, it is not possible to place a limit on the size of deviations from the mass-radius relation ($\epsilon_\text{max}$) or on the coherence length of a background ULDM field ($\Delta x$) because of this ULDM-like signal created by poor measurements of cool WDs.

\indent We conclude that using WDs as precision tests of ULDM is strongly dependent on our understanding of noise sources and observational biases. In our simulation including a background ULDM field, we find that effects from dust and measurement uncertainties are especially important and the most uncertain. Currently, the choices of \texttt{improvement\_factor} and \texttt{effect\_strength} are arbitrary, and future work will need to further characterize the uncertainty in our dust maps and measurements. From our measurements of the spatial correlations in real WD data, we find that observational biases can create ULDM-like signals through a combination of distant-dependent systematic mass-radius offsets and noise. In particular, our measurements of spatial correlations indicate that there are systematic issues with the physical parameters of cool WDs ($T_\text{eff}\lesssim11,000$ K) and that the current measurement technique for spectroscopic surface gravities biases objects towards the peak of the WD mass distribution. Thus, in order to use this method to detect ULDM, better measurements will be needed for these classes of WDs and observational biases will need to be thoroughly understood. Additionally, these signs of issues with particular measurements show that this method can be used to check agreement between theory and measurements as a function of different WD parameters. In the absence of ULDM, we expect all spatial correlations to be noise-dominated, regardless of the WD distance, temperature, radius, surface gravity, or mass. Uncovering positive spatial correlations in these parameters can indicate systematic offsets between theory and measurements.

\indent In this paper, we have demonstrated that, in principle, WDs can be used as tests of ULDM. They can probe both the strength of the coupling and the dark matter mass (from the coherence length). Future work will need to expand our simple simulation to capture new effects not yet addressed here. These expansions could include using a more realistic ULDM field background, employing other approaches to characterizing the spatial correlation such as a graph neural network or using Bayesian methods, and calculating other effects of ULDM on WD structure like variations in the proton and neutron masses, the fine structure constant, and more. Potentially, our technique can be expanded to other types of stars which also obey mass-radius relations to break degeneracies with observational biases \citep{McCrea_1950, Plaut_1953, Huang_1956, Stellar_interiors, Eker_2018}. But, at the current level of measurements, the signal is completely dominated by discrepancies in the mass and radius measurements of low-temperature WDs. 

\section{Acknowledgments} \label{sec:acknow}

\indent N.R.C is supported by the National Science Foundation Graduate Research Fellowship Program under Grant No. DGE2139757. Any opinions, findings, and conclusions or recommendations expressed in this material are those of the author and do not necessarily reflect the views of the National Science Foundation. N.L.Z. acknowledges support by the JHU President’s Frontier Award and by the seed grant from the JHU Institute for Data Intensive Engineering and Science. 

\indent Funding for the Sloan Digital Sky Survey V has been provided by the Alfred P. Sloan Foundation, the Heising-Simons Foundation, the National Science Foundation, and the Participating Institutions. SDSS acknowledges support and resources from the Center for High-Performance Computing at the University of Utah. SDSS telescopes are located at Apache Point Observatory, funded by the Astrophysical Research Consortium and operated by New Mexico State University, and at Las Campanas Observatory, operated by the Carnegie Institution for Science. The SDSS web site is \url{www.sdss.org}.

\indent SDSS is managed by the Astrophysical Research Consortium for the Participating Institutions of the SDSS Collaboration, including the Carnegie Institution for Science, Chilean National Time Allocation Committee (CNTAC) ratified researchers, Caltech, the Gotham Participation Group, Harvard University, Heidelberg University, The Flatiron Institute, The Johns Hopkins University, L'Ecole polytechnique f\'{e}d\'{e}rale de Lausanne (EPFL), Leibniz-Institut f\"{u}r Astrophysik Potsdam (AIP), Max-Planck-Institut f\"{u}r Astronomie (MPIA Heidelberg), Max-Planck-Institut f\"{u}r Extraterrestrische Physik (MPE), Nanjing University, National Astronomical Observatories of China (NAOC), New Mexico State University, The Ohio State University, Pennsylvania State University, Smithsonian Astrophysical Observatory, Space Telescope Science Institute (STScI), the Stellar Astrophysics Participation Group, Universidad Nacional Aut\'{o}noma de M\'{e}xico, University of Arizona, University of Colorado Boulder, University of Illinois at Urbana-Champaign, University of Toronto, University of Utah, University of Virginia, Yale University, and Yunnan University.

\indent This work has made use of data from the European Space Agency (ESA) mission Gaia (\url{https://www. cosmos.esa.int/gaia}), processed by the Gaia Data Processing and Analysis Consortium (DPAC, \url{https://www. cosmos.esa.int/web/gaia/dpac/consortium}). Funding for the DPAC has been provided by national institutions, in particular the institutions participating in the Gaia Multilateral Agreement.

\textit{Software}: astropy \citep{Astropy_2013,Astropy_2018,Astropy_2022}


\bibliographystyle{aasjournal}
\bibliography{bibtex}{}

\end{document}